\documentclass[aps,pra,superscriptaddress,twocolumn,english]{revtex4-2}

\usepackage{graphicx,times}
\usepackage{amsmath,amssymb,amstext}            
\usepackage{graphicx}           
\usepackage{latexsym}
\usepackage{bm}
\usepackage{appendix}
\usepackage{amsmath}
\usepackage[usenames,dvipsnames]{color}
\usepackage{float}
\usepackage{braket}
\usepackage[english]{babel}
\usepackage[colorlinks,bookmarks=false,citecolor=blue,linkcolor=blue,urlcolor=blue]{hyperref}
\newcommand*{\g}{\textsl{g}}
\usepackage[T1]{fontenc}
\usepackage[latin9]{inputenc}
\usepackage{xcolor}
\usepackage{pdfcolmk}
\usepackage{textcomp}
\usepackage{amsmath}
\usepackage{amssymb}
\usepackage{graphicx}
\PassOptionsToPackage{normalem}{ulem}
\usepackage{ulem}

\makeatletter

\providecolor{lyxadded}{rgb}{0,0,1}
\providecolor{lyxdeleted}{rgb}{1,0,0}

\DeclareRobustCommand{\lyxsout}[1]{\ifx\\#1\else\sout{#1}\fi}

\usepackage{braket}
\usepackage{CJK}

\makeatother

\usepackage{babel}

\begin{document}
	
	\title{Collective Effects of Organic Molecules  based on Holstein-Tavis-Cummings Model}
	
	\author{Quansheng Zhang}
	
	\affiliation{Beijing Computational Science Research Center, Beijing 100193, China}
	\email{qshzhang@csrc.ac.cn}
	
	\author{Ke Zhang}
	\affiliation{Helmholtz Institute Mainz, GSI Helmholtzzentrum f$\ddot{u}$r Schwerionenforschung, 55099 Mainz, Germany}
	\affiliation{Johannes Gutenberg-University Mainz, 55099 Mainz, Germany}
	\selectlanguage{english}%
	
	\date{\today}
	\begin{abstract}
	We study the collective effects of an ensemble of organic molecules confined in an optical cavity based on Holstein-Tavis-Cummings model. By using the quantum Langevin approach and adiabatically eliminating the degree of freedom of the vibrational motion, we analytically obtain the expression of the cavity transmission spectrum  to analyze the features of polaritonic states. As an application, we show that the dependence for the frequency shift of the lower polaritonic state on the number of molecules can be used in the detection of the ultra-cold molecules. We also numerically analyze the fluorescence spectrum. The variation of the spectral profile with various numbers of molecules gives signatures for the modification of molecular conformation.
	\end{abstract}
	\maketitle

	\section{Introduction}
	
	The interaction between molecule and light has played an important
	role in photochemical reactivity \cite{Book_Photochememical}, molecular
	spectroscopy \cite{Spec_Plasma_C3CP44103B,Spec_PhysRevLett.118.127401},
	chemical fingerprinting \cite{MFingure_Wrigge2008}, and the generation
	of non-classical states of light \cite{NanClassical_PhysRevLett.97.017402}.
	When molecules located in an optical cavity, the coupling strength
	can be enhanced \cite{Exp_StrongCuplingLidzey1998,Exp_SC_HOLMES200777,ExpSC_PhysRevLett.95.036401,Exp_SC_PhysRevLett.101.116401}.
	In the strong coupling regime, the energy may oscillate between molecules
	and cavity field at a faster rate than their relaxation rates, inducing that
	the molecular and photonic states are hybridized into polaritonic
	states \cite{Rev_doi:10.1063/1.5136320}. In this situation, the potential
	energy landscape for the molecular conformation will be modified,
	giving rise to a range of further effects on the material properties
	of the molecules. As the modifications for molecular conformation
	are tunable, it has opened up new routes for the control of chemical
	reactivity \cite{CavContrR_PhysRevLett.116.238301,RemoteControl_DU20191167},
	energy and charge transport \cite{Excitons_PhysRevLett.114.196403,Charge_PhysRevLett.119.223601},
	and F$\ddot{\text{{o}}}$rster resonance energy transfer \cite{FRET1_https://doi.org/10.1002/anie.201600428,FRET2_Reitz2018,LangveinApproach_PRL}. Besides, the modifications for molecular conformation will bring influence on the profile of  molecular absorption and fluorescence spectra \cite{zeb_exact_2018_ACS}.  For the electronic transitions, the coupling between electronic state and the molecular conformation will lead  to  the structure of the emission (or absorption) spectra observed in experiments \cite{Exp_multi_Klaers2010,Exp_SC_PhysRevLett.101.116401} with multiple peaks.
	The observable molecular spectra will give signatures for the variation
	of molecular conformation. 
	
	To provide an understanding of such phenomena, the Holstein-Tavis-Cummings (HTC) model  has been widely adopted to describe the light-electronic-vibration
	problem \cite{zeb_exact_2018_ACS,Book_charge_nodate}.  {Furthermore, an approach based on the quantum
	Langevin equation  to solve the HTC model has been developed recently
	\cite{LangveinApproach_PRL}. It provides an alternative path to understand the Stokes and anti-Stokes processes \cite{LangveinApproach_PRL}, Purcell effect
	under the  influence of the phononic environments
	\cite{PhysRevResearch.2.033270}, and the Floquent engineering
	\cite{FloquetEngineering_Arxiv} at the level of operators rather than states.
	However, the previous works \cite{LangveinApproach_PRL,PhysRevResearch.2.033270,FloquetEngineering_Arxiv} mainly focus on a single molecule. When multiple
	molecules confined in a cavity, the system will experience more complex
	dynamics. In the case that many molecules are introduced to couple a quantized
	cavity field, an efficient interaction between molecules will be induced,
	leading to collective effects \cite{SERS_Zhang2020}. On the other hand, the description of the
	electronic transition dressed by the vibrational motion suggests that the
	vibrators for one molecule will be coupled to that for other molecules. 
	Considering above facts, a profound influence
	 will be brought on the polaritonic states, as well as the fluorescence
	(or absorption) spectrum. }

	Inspired of this, we provide
	here a further study of HTC model when multiple molecules are 
	presented. Going from a set of coupled
	standard Langevin equations for the whole system, and adiabatically
	eliminating the degree of freedom of the vibrational motion, a set
	{of effective equations can be derived, involving  the average values of Pauli  operators,
	photonic operators and their correlations. From these equations, we study the collective effects in the optical cavity via the cavity transmission spectrum, the steady population and the fluorescence spectrum. {To test the reasonableness of approximation,
    we perform a numerical simulation for the case of two
    molecules in the limit of weak coupling between   the  electronic  states  and molecular conformation with help of the QuTiP package \cite{QuTiP}, and also take perturbative treatment in the first order to compare against  the perturbative method in second order for general situations.} 
    We show that the profiles of cavity transmission  spectrum and the fluorescence spectrum can be modified via the number of molecules, which will give signatures of the modification of molecular conformation. Such phenomena will provide potential applications in the detection of untracold molecules \cite{Detection_PhysRevA.97.063405}, as well as the control of  chemical reaction \cite{RemoteControl_DU20191167,PhysRevLett.106.196405}, energy and charge transport \cite{Charge_PhysRevLett.119.223601,Book_charge_nodate}, and so on.   In contrast, we
	also give the analysis of a special case in the
	limit of without the  coupling between  electronic  states  and
	molecular conformation, i.e. Tavis-Cummings (TC) model.  }
		
	The structure of this paper is organized as following. In Sec.~\ref{sec:Model},
	we describe the model and give the formal solution about the average value
	of the Pauli lowering operator with the adiabatic elimination for
	the degree of freedom vibrational motion. In Sec.~\ref{sec:CT}, we investigate the
	influence of the number of molecules on the polaritonic features via the cavity transmitted field  for
	HTC model and TC model, respectively. In Sec.~\ref{sec:FluorescenceSpectrum},
	we analyze the influence of the number of molecules on the fluorescence
	spectrum  for
	HTC model and TC model, respectively. Finally, a summary is given
	in Sec.~\ref{sec:Conclusion}.
	\section{Model\label{sec:Model}}
	
	\begin{figure}
		\begin{centering}
			\includegraphics[scale=0.45]{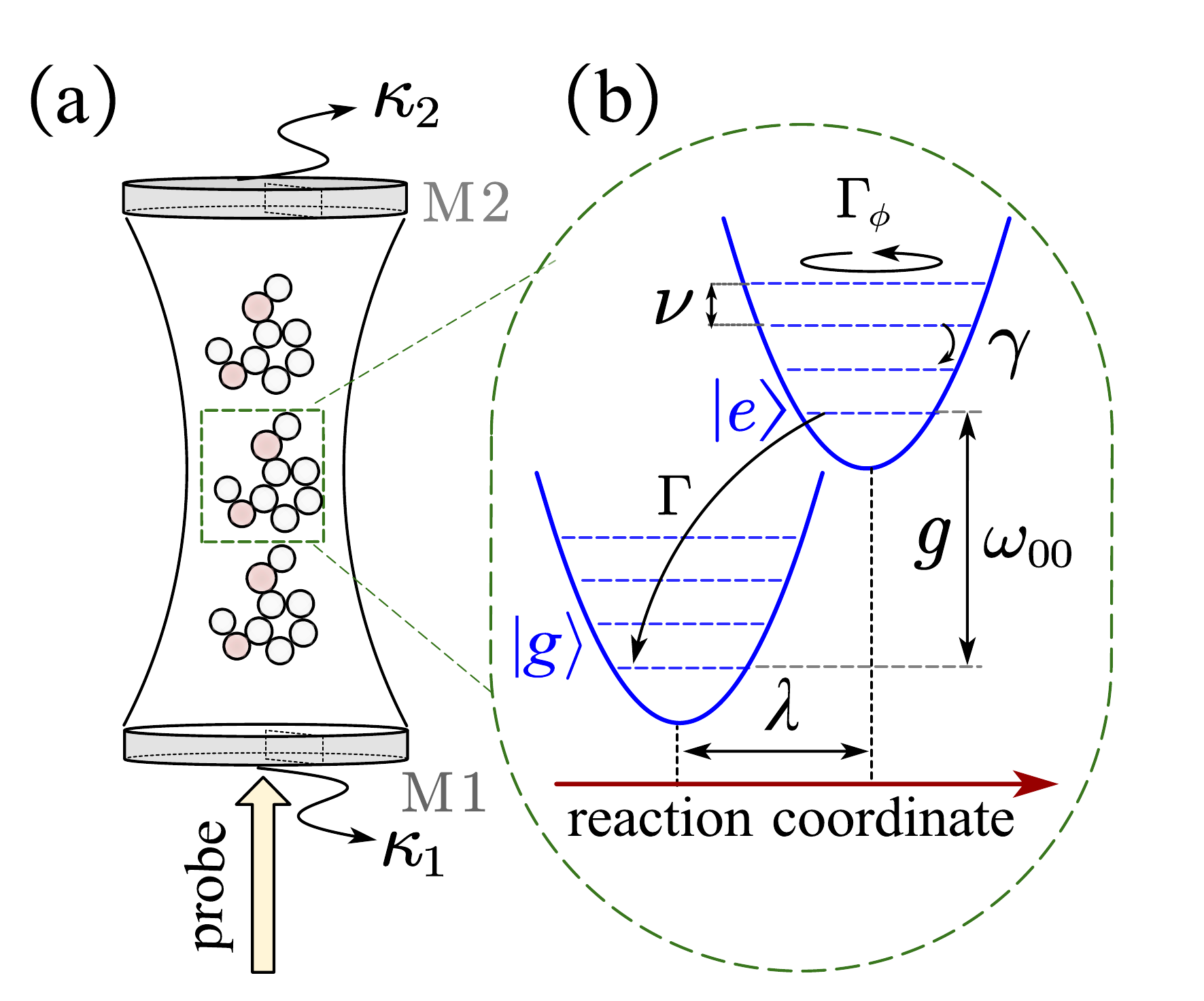}
			\par\end{centering}
		\caption{(Color online) Model schematics. (a) An ensemble of $N$ organic molecules
			are placed inside a Fairy-Perot microcavity with a single mode. The
			dissipation for the cavity field occurs at rates $\kappa_{1}$ through
			the mirror M1 and $\kappa_{2}$ through the mirror M2. (b) Diagram
			of molecular level structure and relaxation processes. Each molecule
			is represented by two harmonic oscillator potential surfaces of two
			electronic states with the shift $\lambda$ in reaction coordinate.
			The cavity field resonantly couples to the electronic transition $\ket{\textsl{g}}\leftrightarrow\ket{e}$
			with the coupling strength $g$. The relaxation processes of each
			molecule include the electronic decay at rate $\Gamma$, and the
			pure dephasing at rate $\Gamma_{\phi}$ and internal vibrational relaxation
			at rate $\gamma$. \label{fig:Model.}}
	\end{figure}
	
	As illustrated in Fig.~\ref{fig:Model.}, the proposed system
of interest consists of an ensemble of $N$ organic molecules, which confined in 
a Fabry-Perot microcavity.  Considering the organic molecules comprised of $z$ atoms, there will be $3z-6$ (or $3z-5$ for linear molecules) normal modes of vibration in the configuration space of the nuclear coordinates ~\cite{Book_charge_nodate,CavContrR_PhysRevLett.116.238301}. When describing a chemical reaction, often a few of these modes, which are associated with the reaction coordinates, are  investigated \cite{JCP_RCHamiltonian,Book_charge_nodate,CavContrR_PhysRevLett.116.238301}.  For a single reaction coordinate, one can assume that the  vibrational frequencies for the ground and excited electronic states are the same. Thus, the system can be represented as a
two-level system (corresponding to its ground $\ket{\g_{m}}$ and
excited $\ket{e_{m}}$ electric states) and a phonon mode (denoting
the harmonic vibrational degree of freedom with the annihilation operator $b_{m}$). \color{black}   Here, each molecule is labeled
by the index $m$ (running from $1$ to $N$).   The dynamic for a single free molecule is then governed by the Holstein Hamiltonian~\cite{Book_charge_nodate}
\begin{equation}
    H_m=\nu b_m^\dagger b_m+[\omega_{e}+\lambda\nu(b_{m}+b_{m}^{\dagger})]\sigma_{m}^{\dagger}\sigma_{m}\label{eq:MoleculeHam},
\end{equation}
where $\omega_{e}=\omega_{00}+\lambda^{2}\nu$ is the effective transition
frequency, with $\omega_{00}$ being the bare splitting between the
ground and excited electric states, and $\nu$ is the vibrational
frequency for the harmonic oscillator. The parameter $\lambda$ is related with the Huang-Rhys factor $S$, i.e. $S=\lambda^{2}$. Here $S$  ranges from $0$ to $\sim2$ \cite{Ultrastrongcoupling_Nanopho,OpticalAbsorptionEmission_OrganicElec,AbsorptionPhotoluminescence_PhysRevA.95.053867}, resulting from the displacement for the equilibrium position of the
vibrational mode between the excited and ground electronic states. Furthermore, $\lambda$ also characters the coupling between the
electronic states and molecular conformation. In the limit of
$\lambda=0$, one can simplify the system for the molecule described by Eq.~(\ref{eq:MoleculeHam})
will to be  a two-level system.
\color{black}

When the cavity field is single-mode, the molecule-cavity
interaction can be depicted by the TC Hamiltonian
(we set $\hbar=1$)
\begin{equation}
H_{\mathrm{int}}=\omega_{c}a^{\dagger}a+g\sum_{m=1}^{N}(a^{\dagger}\sigma_{m}+\mathrm{H.c.}),
\end{equation}
where the operator $a$ annihilates a cavity photon at frequency $\omega_{c}$,
$\sigma_{m}=\ket{\g_{m}}\bra{e_{m}}$ annihilates the molecular excitation
and $g$ is the coupling strength for a single molecule. Notably,
this Hamiltonian implies the assumption that all molecules are equally
coupled to the cavity field via neglecting the influences resulting from
the disorder of transition dipole moment of molecules and molecular spatial
distribution. Such approximation
will conveniently provide a qualitative description of the collective
behavior for organic polaritons \cite{Spano_OrganicCavities}.

Taking account of the free part of the molecules and the situation
that the cavity is probed by a classical field at frequency $\omega_{l}$
with the coupling strength $\eta$, the total Hamiltonian for such
system in the rotating wave approximation can be given 

\begin{equation}
\begin{aligned}H= & \sum_m^N H_m+H_{\mathrm{int}}
  +i\eta(ae^{i\omega_{l}t}-a^{\dagger}e^{-i\omega_{l}t}).
\end{aligned}
\label{eq:Hamiltonian}
\end{equation}

Applying a polaron transformation $H_p=U^\dagger H U$ with 
\begin{equation}
    U=\exp[\sum_m^N\lambda\sigma_m^\dagger\sigma_m(b_m-b_m^\dagger)],
\end{equation}
one can recast the Hamiltonian~(\ref{eq:Hamiltonian}) into
\begin{equation}
\begin{aligned}
H_p=&\sum_m^N(\nu b_m^\dagger b_m+\omega_{00}\sigma_m^\dagger\sigma_m)+\omega_c a^\dagger a\\
&+(\sum_m^N g\mathcal{D}_m^\dagger\sigma_m a^\dagger+i\eta e^{i\omega_l t}a+\mathrm{H.c.}),
\end{aligned}
\end{equation}
where $\mathcal{D}_{m}=\exp[\lambda(b_{m}-b_{m}^{\dagger})]$ is a
displaced operator for $m$-th molecule. In this representation, the coupling between the two-level system and cavity field is dressed by vibrations via  the Frank-Condon factors $_{\mathrm{vib}}\bra{n}\mathcal{D}\ket{m}_{\mathrm{vib}}$ for the transition $\ket{\g}\ket{n}_{\mathrm{vib}}\rightarrow\ket{e}\ket{m}_{\mathrm{vib}}$, where $\ket{n}_{\mathrm{vib}}$ (or $\ket{m}_{\mathrm{vib}}$) are Fock states for vibrations.

\color{black}

To model the open system, we should consider the relaxation processes
as depicted in Fig.~\ref{fig:Model.}(b), including the loss
of cavity field at rate $\kappa=\kappa_{1}+\kappa_{2}$ (encompassing
losses via both mirrors), the spontaneous emission of the  electronic transition
 at rate $\Gamma$,  and the pure dephasing of the electronic
transition at rate $\Gamma_{\phi}$. Their effects are described
as Lindblad terms, and represented by $\mathcal{L}_{\gamma_{\mathcal{O}}}[\mathcal{O}]=\gamma_{\mathcal{O}}(2\mathcal{O}\rho\mathcal{O}^{\dagger}-\mathcal{O}^{\dagger}\mathcal{O}\rho-\rho\mathcal{O}^{\dagger}\mathcal{O})$
with a collapse operator $\mathcal{O}$ and a corresponding decay
rate $\gamma_{\mathcal{O}}$. Finally the dynamics of the system can be described by a master equation of the density
matrix $\rho$
\begin{equation}
\begin{aligned}\dot{\rho}= & -i[H,\rho]+\mathcal{L}_{\kappa}[a]+\sum_{m}^{N}\mathcal{L}_{\Gamma_{\phi}}[\sigma_{m}^{\dagger}\sigma_{m}-\sigma_{m}\sigma_{m}^{\dagger}]\\
 & +\sum_{m}^{N}\mathcal{L}_{\Gamma}[\sigma_{m}]+\sum_{m}^{N}\mathcal{L}_{\gamma}[b_{m}].
\end{aligned}
\label{eq:Master Equation}
\end{equation}
 The last term in Eq.~(\ref{eq:Master Equation}) represents the vibrational
relaxation process at rate $\gamma$. Physically, when considering the coupling
between the vibrator and a reservoir of phonons (also described as Brownian noise dissipation model \cite{LangveinApproach_PRL,PhysRevResearch.2.033270})
in a solvent or surrounding medium, the vibrational relaxation of
molecules cannot be simply expressed in Lindblad form \cite{FloquetEngineering_Arxiv,SpectralOrganicPolaritons_PhysRevA.,PhysRevResearch.2.033270}. {In this work, for the sake of convenience to discuss, we still adopt
the Lindblad decay model to describe the vibrational relaxation process. This
approach has been used in several theoretical studies to explore
the molecular spectroscopy \cite{FloquetEngineering_Arxiv,LangveinApproach_PRL},
Purcell effect \cite{PhysRevResearch.2.033270}, energy transfer \cite{LangveinApproach_PRL}
and organic polariton lasing \cite{OrganicPolaritonLasing_PhysRevLett.}.}  In the limit $\lambda^2\gamma\ll\Gamma$, the Lindblad decay model for the vibrational relaxation will be indistinguishable to the Brownian noise model \cite{FloquetEngineering_Arxiv,LangveinApproach_PRL}. Additionally, the spontaneous emission is also involved in the vibrational motion, which can be ignorable due to its decay rate much smaller
than the vibrational relaxation rate $\gamma$.

The master equation method is usually equivalent to the Langevin approach. For the given master equation in the Lindblad form, described as $\mathcal{L}_{\gamma_{\mathcal{O}}}[\mathcal{O}]$
with a collapse operator $\mathcal{O}$ and decay rate $\gamma_{\mathcal{O}}$, we can map it onto a Langevin form \cite{book_QuantumNoise} as
\begin{equation}
\begin{aligned}\frac{d}{dt}\mathcal{A}= & i[H,\mathcal{A}]-\sum_{j}\left[\mathcal{A},\mathcal{O}_{j}^{\dagger}\right]\left\{ \gamma_{\mathcal{O}_{j}}\mathcal{O}_{j}-\sqrt{2\gamma_{\mathcal{O}_{j}}}\mathcal{O}_{j}^{\mathrm{in}}\right\} \\
 & +\sum_{j}\left\{ \gamma_{\mathcal{O}_{j}}\mathcal{O}_{j}^{\dagger}-\sqrt{2\gamma_{j}}\mathcal{O}_{j}^{\mathrm{in}\dagger}\right\} [\mathcal{A},\mathcal{O}_{j}],
\end{aligned}
\label{eq:Langevin Equation_Form}
\end{equation}
with an arbitrary system operator $\mathcal{A}$ and any collapse
operator $\mathcal{O}_{j}$ in the set \{$b_{m}$, $\sigma_{m}$,
$\sigma_{m}^{\dagger}\sigma_{m}-\sigma_{m}\sigma_{m}^{\dagger}$, $a$\}. 

Under the condition of weak driving $\eta\ll\kappa$,
the cavity field in steady situation will have much less one photon
and the molecules will pretty much stay in the electronic ground state,
i.e. $\sigma_{m}^{\dagger}\sigma_{m}\ll1$. In the rotating frame
at the probe frequency $\omega_{l}$, the quantum Langevin equations
for the cavity field $a$, and polarized operator 
$\tilde{\sigma}_{m}=\mathcal{D}_{m}^{\dagger}\sigma_{m}$
are given as \cite{LangveinApproach_PRL}\begin{subequations}\begin{equation}\begin{aligned}
    \frac{d}{dt}a= & -(i\Delta_{c}+\kappa)a-ig\sum_{m}^{N}\sigma_{m}+\sqrt{2\kappa_{1}}A_{\mathrm{in}} \\
    &+\sqrt{2\kappa_{2}}a_{2}^{\mathrm{in}}(t),\label{eq:Cavity Equation}
    \end{aligned}
\end{equation}
\begin{align}
\frac{d}{dt}\tilde{\sigma}_{m}\approx &
-(i\Delta+\Gamma_{\bot})\tilde{\sigma}_{m}-ig\mathcal{D}_{m}^{\dagger}a\nonumber+\sqrt{2\Gamma_{\bot}}\mathcal{D}_{m}^{\dagger}\sigma_{m}^{\mathrm{in}}
\label{eq:Polarize Operator},\\
\end{align}\label{eq:Cavity-Pauli}\end{subequations}
with the detuning $\Delta_{c}=\omega_{c}-\omega_{l}$
and $\Delta=\omega_{00}-\omega_{l}$.  Here, $\Gamma_{\bot}=\Gamma+2\Gamma_{\phi}$ is the effective transverse relaxation rate. {Notably, under the description of the vibrational relaxation in the Lindblad form, an additional dephasing will be caused to the polarized operator. However, we can neglect such dephasing as this will lead to the disagreement with experimental observations of lifetime for the electronic transition~\cite{LangveinApproach_PRL,FloquetEngineering_Arxiv,PhysRevResearch.2.033270}.}  Additionally, $A_{\mathrm{in}}(t)=\eta/\sqrt{2\kappa_{1}}+a_{1}^{\mathrm{in}}(t)$
denotes the effective cavity input with zero-average input noise $a_{1}^{\mathrm{in}}(t)$. Here we assume that the temperature for cavity field is zero, i.e. $T_{\mathrm{cav}}=0$, then the
non-vanishing correlation $\braket{a_{1}^{\mathrm{in}}(t)a_{1}^{\mathrm{in}\dagger}(t^{\prime})}=\delta(t-t^{\prime})$
and $\braket{a_{2}^{\mathrm{in}}(t)a_{2}^{\mathrm{in}\dagger}(t^{\prime})}=\delta(t-t^{\prime})$
will be obtained. The electronic transition is also affected by a white
nose input $\sigma_{m}^{\mathrm{in}}$ with nonzero correlation $\braket{\sigma_{m}^{\mathrm{in}}(t)\sigma_{n}^{\mathrm{in}\dagger}(t^{\prime})}=\delta_{mn}\delta(t-t^{\prime})$. 

Then we can formally integrate the Eq.~(\ref{eq:Polarize Operator})
to get the solution for $\tilde{\sigma}_{m}$ and subsequently for
Pauli lowering operator\begin{widetext}

\begin{equation}
\begin{aligned}\sigma_{m}(t)= & -\int_{0}^{t}dt_{1}e^{-(i\Delta+\Gamma_{\bot})(t-t_{1})}\mathcal{D}_{m}(t)\mathcal{D}_{m}^{\dagger}(t_{1})[iga(t_{1})-\sqrt{2\Gamma_{\bot}}\sigma_{m}^{in}(t_{1})] +e^{-(i\Delta+\Gamma_{\bot})t}\mathcal{D}_{m}(t)\mathcal{D}_{m}^{\dagger}(0)\sigma_{m}(0).
\end{aligned}
\label{eq:Formula Solution}
\end{equation}
\end{widetext}

In a solvent or surrounding medium, the coupling between the vibrator and a reservoir of phonons will lead to large relaxation rate for vibrational motion (i.e., $\gamma\gg\Gamma_{\bot}$) \cite{PhysRevResearch.2.033270,LargeVRR_JCP,SpectralOrganicPolaritons_PhysRevA.}. Additionally, we also assume that the vibrational relaxation rate much larger than cavity field (i.e., $\gamma\gg\kappa$). Under these conditions, \color{black}  the timescale
of vibrational relaxation will be much shorter than that for the relaxation of electronic state, and also the decay of cavity field. Therefore,
the vibrational state may be considered to be in the steady state~\cite{SpectralOrganicPolaritons_PhysRevA.}. Via treating the vibrations as a Markovian phonon bath~\cite{NJP,SpectralOrganicPolaritons_PhysRevA.}, \color{black} the degrees of freedom for cavity field and
vibrational motion can be separated as 
\begin{equation}
\braket{a(t_{1})\mathcal{D}_{m}(t)\mathcal{D}_{m}^{\dagger}(t_{1})}\approx\braket{a(t_{1})}\braket{\mathcal{D}_{m}(t)\mathcal{D}_{m}^{\dagger}(t_{1})},
\end{equation}
as well as that for Pauli lowering operators and vibrational motion 
\begin{equation}
\braket{\sigma_{n}(t_{1})\mathcal{D}_{m}(t)\mathcal{D}_{m}^{\dagger}(t_{1})}\approx\braket{\sigma_{n}(t_{1})}\braket{\mathcal{D}_{m}(t)\mathcal{D}_{m}^{\dagger}(t_{1})}.
\end{equation}

{If these molecules are initially prepared
in the ground electronic state, and zero photon populated in cavity,
the last term in Eq.~(\ref{eq:Formula Solution}) will be canceled when
we take average over  the cavity field, the vibrational motion, and 
the electronic state. Finally, the average dipole moment
$\braket{\sigma_{m}}$ can be derived 
\begin{equation}
\begin{aligned}\braket{\sigma_{m}}=  -ig\int_{0}^{\infty}dt_{1}\mathcal{\mathcal{F}}_{m}(t-t_{1})\braket{a(t_{1})},
\end{aligned}
\label{eq:Pauli Equation}
\end{equation}
}with the definition $\mathcal{\mathcal{F}}_{m}(t-t_{1})=\Theta(t-t_{1})\exp[-(i\Delta+\Gamma_{\bot})(t-t_{1})]\braket{\mathcal{D}_{m}(t)\mathcal{D}_{m}^{\dagger}(t_{1})}$, where $\Theta(t)$ is the Heaviside step function. The two-time correlation
for the displacement operators $\braket{\mathcal{D}_{m}(t)\mathcal{D}_{m}^{\dagger}(t_{1})}$ can be expressed as (at $t>t_{1}$)
\begin{equation}
\braket{\mathcal{D}_{m}(t)\mathcal{D}_{m}^{\dagger}(t_{1})}=e^{-\lambda^{2}}e^{\lambda^{2}e^{-(\gamma+i\nu)(t-t_{1})}},\label{eq:DD CORR}
\end{equation}
when the effective vibrational temperature satisfies $k_{B}T_{\mathrm{vib}}/\hbar\nu\ll1$
(see Appendix~\ref{sec:DynamicVibration}). The important quantity to be emphasized is the function $\braket{\mathcal{D}_{m}(t)\mathcal{D}_{m}^{\dagger}(t_{1})}$, which implies the influence of the vibrational mode on the excitons
in organic molecules. When $\lambda=0$, one can achieve the value of the function $\braket{\mathcal{D}_{m}(t)\mathcal{D}_{m}^{\dagger}(t_{1})}=1$.
Then the interaction between electronic states and molecular conformation
will be removed. For finite $\lambda$, the coupling between the vibrational mode and the excitons will result in some new fascinating physics.

Notably, the Pauli operator can  feed back into itself via the cavity field, resulting in higher order correlation between the displaced operators. The adiabatic approximation  depicted by Eq.~(\ref{eq:DD CORR}) has neglected the contributions from these higher order correlations. To explore the influence of these higher order correlations, we have taken further discussion in Appendix~\ref{sec:Dynamic_DO} and \ref{sec: Appendix Steady State}.
\color{black}

	
	\section{Cavity Transmission \label{sec:CT}}
	
	 \begin{figure*}[htp]
		\centering{}\centering\includegraphics[width=0.95\linewidth]{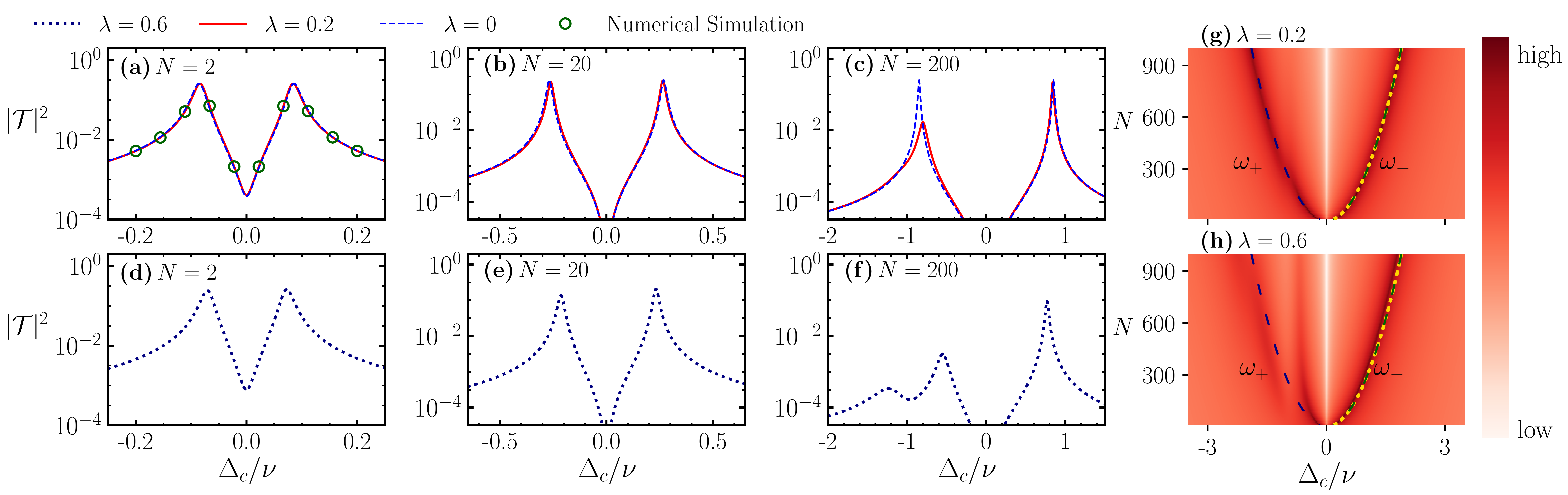}\caption{(Color online) Cavity-molecule spectroscopy. {(a-f) Influence of the number $N$ (increases as $2$, $20$, and $200$) of  identical organic molecules  on  the  intensity of transmitted field at resonance $\omega_{c}=\omega_{00}$   at $\lambda=0$ (blue dashed line), $\lambda=0.2$ (red solid line) and $\lambda=0.6$ (black dotted line), respectively. (a) The green circle marks shows the numerical simulation results for the case $\lambda=0.2$, $N=2$ with the help of QuTiP.  Intensity of transmitted field
			at resonance $\omega_{c}=\omega_{00}$ as a function of the detuning
			$\Delta_{c}/\nu$ and the number $N$ of  organic molecules  at $\lambda=0.2$ (g) and $\lambda=0.6$ (h), respectively. The spatial
			intensity profile of the transmitted field is shown in arbitrary units
			but to scale. The green and  blue  dashed lines show the 	frequency shift of the lower and upper polaritonic states various the number $N$ of
			organic molecules, respectively, given by Eq.~(\ref{eq:UP-LP_Fre}).  The frequency shift of the lower polaritonic state is about $\omega_{-}\approx g\sqrt{N}$ (yellow dotted line).  The other parameters are $\nu/\Gamma=250$,
			$\Gamma_{\phi}/\Gamma=2\kappa_{1}/\Gamma=2\kappa_{2}/\Gamma=1$, $\gamma/\Gamma=50$,
			$g/\Gamma=5$, and $\eta/\Gamma=0.01$.}\label{fig:CavitySpec}}
	\end{figure*}

		For the molecule-cavity system discussed in the previous section,
	the electronic transition will be dressed by vibrational
	motion with the coupling between the electronic states and molecular conformation.
	In the limit of large vibrational relaxation (i.e. $\gamma\gg\kappa$
	and $\gamma\gg\Gamma_{\bot}$), the vibrational mode can be treated as a locale
	phonon reservoir under the Markov approximation. Going from the Langevin
	equation of polarized operator and adiabatically eliminating the
	degree of freedom of vibrational motion, we formally derive the expression
	of \textcolor{black}{the average Pauli lowering operator $\braket{\sigma_{m}}$}.
	It is noteworthy that the first term in Eq.~(\ref{eq:Pauli Equation})
	represents a convolution. This character will inspire us to utilize the Laplace transformation (defined as $\braket{\bar{f}}(s)=\int_{0}^{\infty}dt\:f(t)e^{-st}$
	for time dependent function $f(t)$ at $t\geq0$) to   simplify the
	derivation. Then, Eq.~(\ref{eq:Pauli Equation}) can be reformed as
	\begin{equation}
	\braket{\bar{\sigma}_{m}}=-ig\bar{\mathcal{F}}_{m}(s)\braket{\bar{a}},\label{eq:Pauli LapLace Transform}
	\end{equation}
	where $\bar{\mathcal{F}}_{m}$ is the Laplace transform of $\mathcal{\mathcal{F}}_{m}(t)$, expressed as
	\begin{equation}
	\bar{\mathcal{F}}_{m}=\sum_{k}\frac{\lambda^{2k}}{k!}\frac{e^{-\lambda^{2}}}{s+i(\Delta+k\nu)+(\Gamma_{\bot}+k\gamma)},
	\end{equation}
	with the Laplace transform
	variable $s$. Similarly, taking average over the cavity field as well
	as \textcolor{black}{the electronic degrees of freedom on both sides
		of} Eq.~(\ref{eq:Cavity Equation}), and applying the Laplace transformation,
	we finally obtain
	\begin{equation}
	s\braket{\bar{a}}=-(i\Delta_{c}+\kappa)\braket{\bar{a}}-ig\sum_{m}^{N}\braket{\bar{\sigma}_{m}}+\frac{\eta}{s}.\label{eq:Cavity Laplace}
	\end{equation}

	Considering the system with $N$ identical molecules, the Laplace
	form for the average Pauli lowering operator $\braket{\bar{\sigma}_{m}}$
	and the function $\bar{\mathcal{F}}_{m}(s)$ would be the same for
	any molecule ($m$). Combining the Laplace forms of the Pauli operator described	by Eq.~(\ref{eq:Pauli LapLace Transform}) and the cavity field illustrated by Eq.~(\ref{eq:Cavity Laplace}), we can get \begin{subequations}
		\begin{align}
		\braket{\bar{\sigma}_{m}} & =-\frac{ig\eta\bar{\mathcal{F}}_{m}}{s[Ng^{2}\bar{\mathcal{F}}_{m}+i\Delta_{c}+\kappa+s]},\\
		\braket{\bar{a}} & =\frac{\eta}{s[i\Delta_{c}+s+\kappa+Ng^{2}\bar{\mathcal{F}}_{m}]}.
		\end{align}
	\end{subequations} 
	From the final value theorem \cite{SpataniousEmssion&Laplace_PhysRevA.74.033816},
	we get the steady values\begin{subequations}\begin{align}
	 \braket{\sigma_m}_{\mathrm{ss}}&=\lim_{s\rightarrow0}s\braket{\bar{\sigma}_m}=-\frac{ig\eta\chi}{i\Delta_{c}+\kappa+Ng^{2}\chi}\label{eq:Pauli Steady},\\
	\braket{a}_{\mathrm{ss}}&=\lim_{s\rightarrow0}s\braket{\bar{a}}=\frac{\eta}{(i\Delta_{c}+\kappa)+Ng^{2}\chi}\label{eq:Cavity Steady},
	\end{align}\label{eq:Steady}\end{subequations}where ``ss''
	stands for the steady state situation, and $\chi=\underset{s\rightarrow0}{\lim}\bar{\mathcal{F}}_{m}$. The item
	$Ng^{2}\chi$ in Eq.~(\ref{eq:Cavity Steady})
	reflects the hybridization between photonic states and molecule electronic
	states. One can approximately obtain the hybridized decay rates and
	frequencies (under resonance conditions, i.e., fixed $\Delta=\Delta_{c}=0$)
	\cite{PurcellEffect_PhysRevA.99.043843,Rev_doi:10.1063/1.5136320} \begin{subequations}
		\begin{align}
		\Gamma_{\pm} & =\frac{\Gamma_{\mathrm{eff}}+\kappa}{2}\pm\mathfrak{R}\sqrt{\frac{(\Gamma_{\mathrm{eff}}+i\Delta_{\mathrm{eff}}-\kappa)^{2}}{4}-Ng^{2}},\label{eq:UP_LPDecay}\\
		\omega_{\pm} & =-\frac{\Delta{}_{\mathrm{eff}}}{2}\pm\mathfrak{I}\sqrt{\frac{(\Gamma_{\mathrm{eff}}+i\Delta_{\mathrm{eff}}-\kappa)^{2}}{4}-Ng^{2}},\label{eq:UP-LP_Fre}
		\end{align}
	\end{subequations}
	where $\Gamma_{\mathrm{eff}}=\mathfrak{R}{\lim}1/\chi$
	and $\Delta_{\mathrm{eff}}=\mathfrak{I}{\lim}1/\chi$
	denote the effective decay rate and additional frequency shift for
	excitons, respectively.
	
	In experiments, the hybridized decay rates and frequencies can be
	detected via the cavity transmitted field. According to the input
	and output relation $a_{2,\mathrm{out}}(t)=\sqrt{2\kappa_{2}}a(t)$,
	the transmission $\mathcal{T}$ via $\mathcal{T}=\braket{a_{2,\mathrm{out}}}_{\mathrm{ss}}/\braket{A_{\mathrm{in}}}_{\mathrm{ss}}$
	in steady state reign will be expressed as
	\begin{equation}
	\mathcal{T}=\frac{2\sqrt{k_{1}k_{2}}}{(i\Delta_{c}+\kappa)+Ng^{2}\chi}.
	\end{equation}

	Figures~\ref{fig:CavitySpec}(a-f) plot the intensity of the cavity
	transmission at resonance $\omega_{c}=\omega_{00}$ in the various
	number $N$ of the organic molecules   for TC (see blue dashed line) and HTC (see red
	solid line and black dotted line) model, respectively. The parameters we chosen are given
	in the caption of Fig.~\ref{fig:CavitySpec}, which  associate
	with the recent theoretical researches \cite{LangveinApproach_PRL,PhysRevResearch.2.033270,OrganicPolaritonLasing_PhysRevLett.,FloquetEngineering_Arxiv}.
	\textcolor{black}{When multiple molecules emerged in a cavity, the system
		will be in the collective behavior with a bright mode that couples
		to the cavity field with an enhanced strength $g\sqrt{N}$ (for identical
		molecules). }In the limit of $\lambda=0$,  {[}see
	blue dashed line in Fig.~\ref{fig:CavitySpec}(a-c){]}, the cavity
	transmission will be split into two symmetric peaks with the central frequency
	separation $2\sqrt{Ng^{2}-(\Gamma_{\bot}+ \kappa)^{2}/4}$ \cite{Bernardot_1992_VRS_Detection_PRA,khitrova_VRS_Nature,PurcellEffect_PhysRevA.99.043843}.
	
	For finite $\lambda$, the coupling between electronic states and
	molecular conformation will produce additional dephasing as well as
	the frequency shift for excitons, resulting in an asymmetric cavity
	transmission profile {[}see red solid line in Fig.~\ref{fig:CavitySpec}(c) and black dotted line in Figs.~\ref{fig:CavitySpec}(d-f){]}. {In the limit of $\lambda^2\ll1$ and $g\sqrt{N}\ll\nu$, the evolution of the transmitted field for HTC model with the detuning $\Delta_c$ will perform  similar behaviors to the TC model, due to the small Franck-Condon overlap and the weak collective effects.} To clarify the influence of the number $N$ of organic
	molecules on the hybridized frequencies as well as the decay rates,
	we  numerically show the spatial intensity profile of the transmitted
	field. It is clearly to see that the evolution of the frequency for lower polariton {with the number $N$ of molecules obeys the rule given by Eq.~(\ref{eq:UP-LP_Fre}) [see green dashed line in Fig.~\ref{fig:CavitySpec}(g, h)]}. For blue-detuned illumination,
	the spatial intensity profile of the transmitted field seems more
	complicated {[}see Fig.~\ref{fig:CavitySpec}(c-h){]}. Typically, when
	the collective coupling strength $g\sqrt{N}>\nu$ for large Franck-Condon overlap (i.e. $\lambda^2\sim1$), one more polaritonic
	peak appears in the cavity transmission profile {[}see black dotted line Fig.~\ref{fig:CavitySpec}(f){]}.
	For the left branch in Fig.~\ref{fig:CavitySpec}(h), the relation between the number $N$ of
	molecules and the central frequency still obeys the rule given
	in Eq.~(\ref{eq:UP-LP_Fre}) {[}see  blue dashed line in Fig.~(\ref{fig:CavitySpec})(h){]}.
	This  branch corresponds to the upper polaritons. The appearance of
	the another branch is caused by the perturbation for the excitation
	state of the vibrational motion, corresponding to the dark polaritonic state
	\cite{Rev_doi:10.1063/1.5136320}.  
	
	{In addition, we would like to point out here the cavity transmitted field can be used to detect the ultra-cold molecules \cite{Detection_PhysRevA.97.063405}. Considering the presence of rovibrational states, several theoretical studies have modeled one molecule as a several-level system \cite{CollectiveDissipative_PhysRevLett.125.193201,Detection_PhysRevA.97.063405}.  One can estimate the number of  molecules  via the frequency separation between upper and lower polaritonic states. In contrast to the previous work \cite{Detection_PhysRevA.97.063405}, the  coupling  between  electronic  states  and molecular conformation described by HTC model will make the cavity transmitted field become more complicated [see Fig. \ref{fig:CavitySpec}(g-h)]. Particularly for the large Franck-Conda overlap,  it will be a challenge task to detect molecules according to the method introduced by Ref. \cite{Detection_PhysRevA.97.063405} due to the appearance of the peak for dark polaritonic state, and the suppressed intensity for the upper polaritonic state. Notably, a single peak appeared in the spatial intensity profile of the transmitted field when $\Delta_c>0$, which is associated with the lower polaritons. Considering the relation between the number $N$ of  molecules and the central frequency of the lower polaritonic peaks described by Eq.~(\ref{eq:UP-LP_Fre}), we can approximately get the number of  molecules  as $N\approx \omega_{-}^2/g^2$ [see yellow dotted line in Fig. \ref{fig:CavitySpec}(g, h)].}

	
	\section{Fluorescence Spectrum\label{sec:FluorescenceSpectrum}}
	In the previous section, we have analyzed the polaritonic features via 
	the transmission field profile.  With the increasing
	of the number of molecules, qualitatively different phenomena arise while considering
	the collective behavior for the emitters. On the other hand, the molecules can interact with each other via the
	coupling with the quantized cavity field. This will give rise to the modification of the molecular
	configuration \cite{CBO_PhysRevX.9.021057}, as well as the transition between the electronic
	ground and excited states. In this section, we will illustrate how
	the feature is reflected via the fluorescence spectrum. 
	
	Before the introduction of fluorescence spectrum, let us first give the
	definition of the fluctuation operator. Since the expect values for
	Pauli lowering operator $\braket{\sigma_{m}}_{\mathrm{ss}}$ and the
	cavity field $\braket{a}_{\mathrm{ss}}$ have been obtained in steady
	state  given by Eq.~(\ref{eq:Steady}),
	one can represent the fluctuation about this steady state by the
	following operators
	\begin{equation}
	\begin{aligned}\delta\sigma_{m} & =\sigma_{m}-\braket{\sigma_{m}}_{\mathrm{ss}}\\
	\delta a & =a-\braket{a}_{\mathrm{ss}}.
	\end{aligned}
	,\label{eq:Fluctuation operator}
	\end{equation}
	Considering the formal solution for Pauli operator given by Eq.~(\ref{eq:Formula Solution}),
	one can easily derive the expression for the autocorrelation function
	\begin{widetext} 
	\begin{align}
	    \braket{\delta\sigma_{m}^{\dagger}(t)\delta\sigma_{m}(t+\tau)}=&	-ig\int_{t}^{t+\tau}dt_{1}e^{-(i\Delta+\Gamma_{\bot})(t+\tau-t_{1})}\braket{\delta\sigma_{m}^{\dagger}(t)\mathcal{D}_{m}(t+\tau)\mathcal{D}_{m}^{\dagger}(t_{1})\braket{a}_{\mathrm{ss}}}-\braket{\sigma_{m}}_{\mathrm{ss}}\braket{\delta\sigma_{m}^{\dagger}(t)}\nonumber\\
	&-ig\int_{t}^{t+\tau}dt_{1}e^{-(i\Delta+\Gamma_{\bot})(t+\tau-t_{1})}\braket{\delta\sigma_{m}^{\dagger}(t)\mathcal{D}_{m}(t+\tau)\mathcal{D}_{m}^{\dagger}(t_{1})\delta a(t_{1})}\label{eq:Auto_Correlation}\\
	&+e^{-(i\Delta+\Gamma_{\bot})\tau}\braket{\delta\sigma_{m}^{\dagger}(t)\mathcal{D}_{m}(t+\tau)\mathcal{D}_{m}^{\dagger}(t)\sigma_{m}(t)}.\nonumber
	\end{align}
		Under the large vibrational relaxation condition $\gamma\gg\kappa$
		and $\gamma\gg\Gamma$, the correlation time for vibrational motion
		would be much shorter than the timescale of the correlation between
		cavity field and molecules, as well as the intra-molecule correlations.
		Therefore, the four-operator correlation functions can be separated
		as 
		\begin{equation}
		\begin{aligned}\braket{\delta\sigma_{m}^{\dagger}(t)\mathcal{D}_{m}(t+\tau)\mathcal{D}_{m}^{\dagger}(t_{1})\mathcal{O}(t_{1})} & \approx\braket{\delta\sigma_{m}^{\dagger}(t)\mathcal{O}(t_{1})}\braket{\mathcal{D}_{m}(t+\tau)\mathcal{D}_{m}^{\dagger}(t_{1})}\end{aligned}
		\label{eq:FourTo2Two}
		\end{equation}
		with $\mathcal{O}(t)$ in the set $\{$$a(t)$, $\sigma_{m}(t)$,
		$\sigma_{n}(t)$$\}$. In the limit of a long time scale, the system would be
		in the steady state. Then the first line in Eq.~(\ref{eq:Auto_Correlation})
		can be neglected as the expectation value for fluctuation operator $\braket{\delta\sigma_{m}}_{\mathrm{ss}}$
		will be zero. The formulation for the desired
		auto-correlation function is deduced with convolution integral
		\begin{equation}
		\braket{\delta\sigma_{m}^{\dagger}(0)\delta\sigma_{m}(\tau)}_{\mathrm{ss}}=-ig\int_{0}^{\infty}dt_{1}\mathcal{F}_{m}(\tau-t_{1})\braket{\delta\sigma_{m}^{\dagger}(0)\delta a(t_{1})}_{\mathrm{ss}}-\braket{\delta\sigma_{m}^{\dagger}\delta\sigma_{m}}_{\mathrm{ss}}\mathcal{F}_{m}(\tau),\label{eq:TwoTimeCorr}
		\end{equation}
	\end{widetext} where $\braket{\delta\sigma_{m}^{\dagger}(0)\delta\mathcal{O}(\tau)}_{\mathrm{ss}}\equiv\underset{t\rightarrow\infty}{\lim}\braket{\delta\sigma_{m}^{\dagger}(t)\delta\mathcal{O}(t+\tau)}_{\mathrm{ss}}$
	and $\braket{\delta\sigma_{m}^{\dagger}\delta\sigma_{m}}_{\mathrm{ss}}=\braket{\sigma_{m}^{\dagger}\delta\sigma_{m}}_{\mathrm{ss}}-\braket{\delta\sigma_{m}}_{\mathrm{ss}}\braket{\sigma_{m}^{\dagger}}_{\mathrm{ss}}$.
	Taking the Fourier transformation with the definition $S_{\mathcal{O}}^{m}(\omega)=\int_{-\infty}^{\infty}d\tau\,\braket{\delta\sigma_{m}^{\dagger}(0)\delta\mathcal{O}(\tau)}_{\mathrm{ss}}e^{-i\omega\tau}$,
	the correlation function given by Eq.~(\ref{eq:TwoTimeCorr}) can
	be reformed in the Fourier domain, i.e., 
	\begin{equation}
	S_{\sigma_{m}}^{m}=-\mathcal{\tilde{F}}_{m}[igS_{a}^{m}-\braket{\delta\sigma_{m}^{\dagger}\delta\sigma_{m}}_{\mathrm{ss}}],\label{eq:Spec Mth Molecule}
	\end{equation}
	where $\mathcal{\tilde{F}}_{m}(\omega)$ is the Fourier form for the
	factor $\mathcal{F}_{m}(t)$. 
	
	Taking the similar method, one can respectively  get the Fourier
	forms for the correlation functions $\braket{\delta\sigma_{m}^{\dagger}(0)\delta\sigma_{n}(\tau)}_{\mathrm{ss}}$ between molecules,
	 and also the correlation function $\braket{\delta\sigma_{m}^{\dagger}(0)\delta a(\tau)}_{\mathrm{ss}}$ between cavity and molecules
	\begin{align}
	\begin{aligned}S_{\sigma_{n}}^{m}= & -\mathcal{\tilde{F}}_{n}(igS_{a}^{m}-\braket{\delta\sigma_{m}^{\dagger}\delta\sigma_{n}}_{\mathrm{ss}}),\\
	S_{a}^{m}= & -ig\tilde{\mathcal{F}}_{a}(\sum_{n\neq m}^{N}S_{\sigma_{n}}^{m}+S_{\sigma_{m}}^{m})+\tilde{\mathcal{F}}_{a}\braket{\delta\sigma_{m}^{\dagger}\delta a}_{\mathrm{ss}},
	\end{aligned}
	\label{eq:Spec M-N Pair}
	\end{align}
	where $\braket{\delta\sigma_{m}^{\dagger}\delta\mathcal{O}}_{\mathrm{ss}}=\braket{\sigma_{m}^{\dagger}\mathcal{O}}_{\mathrm{ss}}-\braket{\mathcal{O}}_{\mathrm{ss}}\braket{\sigma_{m}^{\dagger}}_{\mathrm{ss}}$,
	and $\tilde{\mathcal{F}}_{a}(\omega)=1/[i(\Delta_{c}+\omega)+\kappa]$. 
	
	For identical molecules, the functions $S_{\sigma_{n}}^{m}$ ($n\neq m$)
	are same for any molecule pair ($m$, $n$) and $S_{\sigma_{m}}^{m}$
	are interchangeable for any molecule ($m$). As a result, Eqs.~(\ref{eq:Spec Mth Molecule}-\ref{eq:Spec M-N Pair})
	can be simply reformed as 
	\begin{equation}
	\mathcal{M}_{s}\boldsymbol{\mathcal{S}}_\mathrm{Vec}+\boldsymbol{\mathcal{S}}_{\mathrm{in}}=0,
	\end{equation}
	with the coefficient matrix
	\[
	\mathcal{M}_{s}=\left(\begin{array}{ccc}
	-1 & 0 & -ig\mathcal{\tilde{F}}_{m}\\
	0 & -1 & -ig\mathcal{\tilde{F}}_{n}\\
	-ig\mathcal{\tilde{F}}_{a} & -i(N-1)g\mathcal{\tilde{F}}_{a} & -1
	\end{array}\right),
	\]
	the vector for the Fourier form of correlation functions $\boldsymbol{\mathcal{S}}_\mathrm{Vec}=($$S_{\sigma_{m}}^{m},$
	$S_{\sigma_{n}}^{m}$, $S_{a}^{m})^{T}$, and the input vector $\boldsymbol{\mathcal{S}}_{\mathrm{in}}=($$\mathcal{\tilde{F}}_{m}\braket{\delta\sigma_{m}^{\dagger}\delta\sigma_{m}}_{\mathrm{ss}}$,
	$\mathcal{\tilde{F}}_{n}\braket{\delta\sigma_{m}^{\dagger}\delta\sigma_{n}}_{\mathrm{ss}}$,
	$\mathcal{\tilde{F}}_{a}\braket{\delta\sigma_{m}^{\dagger}\delta a}_{\mathrm{ss}})^{T}$.
	To this end, the solution can be obtained via $\boldsymbol{\mathcal{S}}_\mathrm{Vec}=-\mathcal{M}_{s}^{-1}\boldsymbol{\mathcal{S}}_{\mathrm{in}}.$
	It should be noted that the above equation can also be derived via
	considering the quantum regression theorem. Different with the traditional
	methods \cite{Book_carmichael_statistical_1999}, the convolution integral for $\braket{\sigma_{m}(t)}$
	given by Eq.~(\ref{eq:Pauli Equation}) requires us to get the dynamic
	equation for the correlation function from the perspective of the
	frequency domain. In addition, to obtain the value of the vector $\boldsymbol{\mathcal{S}}_\mathrm{Vec}$,
	one need to calculate the expectations $\braket{\sigma_{m}^{\dagger}\mathcal{O}}$
	in steady state, which have been discussed in Appendix~\ref{sec: Appendix Steady State}.
	
	According to the definition of the collective operator $J_{-}=\sum_{m=0}^{N}\sigma_{m}$,
	the fluorescence spectrum in steady situation can be achieved by taking
	the Fourier transformation $S=\underset{t\rightarrow\infty}{\mathrm{lim}}\mathfrak{R}\int_{-\infty}^{\infty}d\tau\,\braket{\delta J_{-}^{\dagger}(t)\delta J_{-}(t+\tau)}e^{-i\omega\tau}$.
	Since $\braket{\delta\sigma_{m}^{\dagger}(t)\delta\sigma_{m}(t+\tau)}$,
	$\braket{\delta\sigma_{m}^{\dagger}(t)\delta\sigma_{n}(t+\tau)}$
	($m\neq n$) are same for all molecules and all molecular
	pairs respectively, we can further get the relation
	\begin{equation}
	S=N\mathfrak{R}S_{\sigma_{m}}^{m}+N(N-1)\mathfrak{R}S_{\sigma_{n}}^{m}.
	\end{equation}
	
	
	\begin{figure}
		\centering{}\includegraphics[scale=0.24]{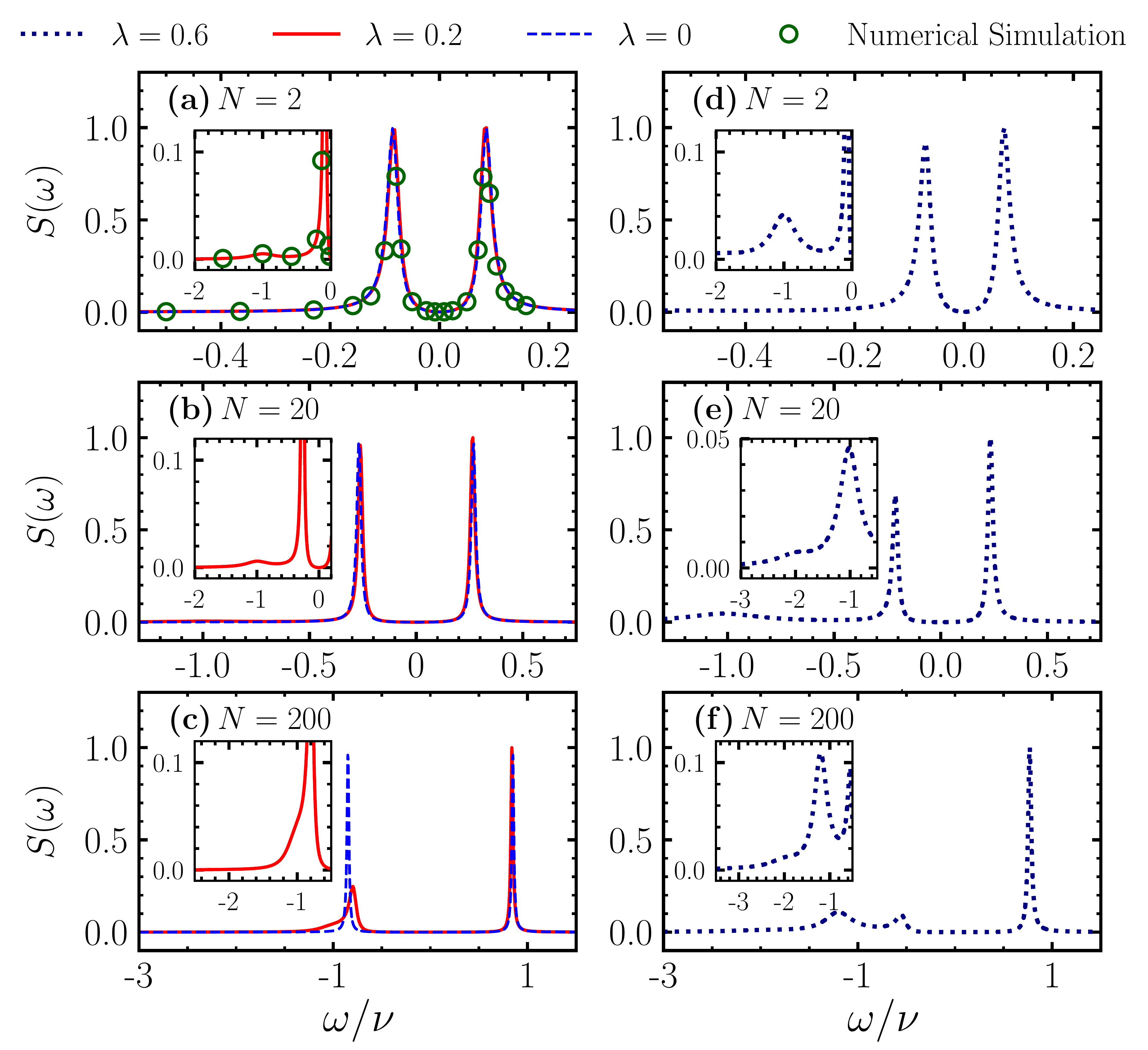}\caption{(Colored online) The fluorescence spectrum at resonance $\omega_{c}=\omega_{00}=\omega_{l}$
			with various number $N$ of organic molecules  (increases as $2$, $20$, and $200$). We compare the HTC model  for different number $N$ of molecules, against the TC model
			(blue dashed line) when (a-c) $\lambda=0.2$  and (d-f) $\lambda=0.6$. (a) The green circle markers are numerical calculated with the help of QuTiP
			for the case $\lambda=0.2$, and $N=2$. The
			other parameters are the same with Fig.~\ref{fig:CavitySpec}. \label{fig:FluorescenceSpectrum}}
	\end{figure}
	
	
	Figure~\ref{fig:FluorescenceSpectrum} illustrates the fluorescence
	spectrum at resonance $\omega_{c}=\omega_{00}=\omega_{l}$. With the number $N$ of molecules increasing, the spectral weight shows some interesting
	features. In the limit of $\lambda=0$, the profile for the fluorescence
	spectrum will exhibit two symmetric peaks with the central frequencies
	$\omega_{\pm}$ given by Eq.~(\ref{eq:UP-LP_Fre}) (see blue dashed
	line). For finite $\lambda$, the coupling between electronic and
	molecular conformation will bring us more amazing phenomena, e.g.
	the Stokes scatting, the asymmetric profile for fluorescence spectrum
	[see red solid line in Fig.~\ref{fig:FluorescenceSpectrum}(a-c) and black dotted line in Fig.~\ref{fig:FluorescenceSpectrum}(d-f)]. When $N=2$, one more peak at frequency $\omega=-\nu$ appeared  compared with TC  (i.e. $\lambda=0$) model {[}see red
	solid line in Fig.~\ref{fig:FluorescenceSpectrum}(a) and black dotted line in Fig.~\ref{fig:FluorescenceSpectrum}(d){]}. This
	peak corresponds to the Stokes process  $\ket{e}\ket{0}_{\mathrm{vib}}\rightarrow\ket{\g}\ket{1}_{\mathrm{vib}}$,
	with the transition rate characterized by the Franck-Condon factors $_{\mathrm{vib}}\bra{1}\mathcal{D}(\lambda)\ket{0}_{\mathrm{vib}}$
	\cite{Fluorensence_sauer_handbook_2011}. 
	As the number $N$  of organic molecules increases, the coverage of the
	Stokes line in the frequency domain will be broadened [see black dotted line in Fig.~\ref{fig:FluorescenceSpectrum}(e, f)].  Especially, at $N=20$,
	one can find that another peak appeared at frequency $\omega=-2\nu$,
	corresponding to the Stoke process $\ket{e}\ket{0}_{\mathrm{vib}}\rightarrow\ket{\g}\ket{2}_{\mathrm{vib}}$.
	The dependence of the profile for the Stokes line on the number of
	molecules reflects that the molecular conformation has been modified
	via the collective behavior. When the collective coupling strength
	$g\sqrt{N}>\nu$, the contribution of dark polartonic state should
	be taken into account {[}see black dotted line in Fig. \ref{fig:FluorescenceSpectrum}(f){]},
	which would bring new physics. 
	
	\section{Conclusion\label{sec:Conclusion}}
	In summary, we have studied the collective behavior for organic molecules
	based on the HTC model. Via adiabatically eliminating the degree of
	freedom of the vibrational motion, we have derived a set of linear
	equations in Laplace (or Fourier) domain, involving the
	Pauli operators, the correlation between cavity field and molecules,
	and the correlation between molecules. Based on these equations,
	we have studied the cavity transmission spectrum for an ensemble
	of identical organic molecules.   Our results reveal that, different with  TC model, the coupling between electronic states and molecular
	conformation will produce additional dephasing as well as the
	frequency shift for excitons. This will result in some amazing
	phenomena, e.g. the suppressed upper polaritonic peak,
	dark polaritonic peak with central frequency slightly changed via
	increasing the number of molecules. In addition, the cavity
	transmitted field is suggested to detect the ultra-cold
	molecules \cite{Detection_PhysRevA.97.063405}. Considering the relationship between the hybridized frequencies and the number of molecules, one can estimate the number of molecules  via the 
	frequency shift for the lower polaritonic state. Moreover, we also study the steady 	population on the excited state for  total molecules   and the fluorescence spectrum. With the number of molecules increasing, the coverage of the Stokes line in the frequency domain is broadened.  This means that the probabilities for transition
	from electronic excited state to the vibrational excited states  in the
	electronic ground state increased. In other words, the conformation of molecules will be modified due to the collective effects, providing a potential application in the control of the chemical
	reactivity \cite{CavContrR_PhysRevLett.116.238301,RemoteControl_DU20191167},
	energy and charge transport \cite{Excitons_PhysRevLett.114.196403,Charge_PhysRevLett.119.223601},
	and F$\ddot{\text{{o}}}$rster resonance energy transfer \cite{FRET1_https://doi.org/10.1002/anie.201600428,FRET2_Reitz2018,LangveinApproach_PRL}.

	\begin{acknowledgments}
		This work was supported by the Natural Science Foundation of China (under
		Grants No. 12074030 and No. U1930402). Q. Z. acknowledges fruitful discussions with Lei Du, Yong Li and Michael Reitz.
	\end{acknowledgments}
	
	
	\appendix
	
	\section{Dynamic of the Vibrational Motion\label{sec:DynamicVibration}}
	In the limit of weak probe fields $\eta\ll\kappa$, the Heisenberg-Langevin equation for the vibrational motion is given  \cite{LangveinApproach_PRL} by
	\begin{equation}
	    \frac{d}{dt}b=-i(\nu-i\gamma)b+\sqrt{2\gamma}b_{\mathrm{in}},\label{eq:Ossilator}
	\end{equation}
	where $b_{\mathrm{in}}$ is the annihilation operator of thermal noise  with zero-average value. The non-vanishing correlation functions are
	\begin{align}
	   \begin{aligned}
	   \braket{b_{\mathrm{in}}^{\dagger}(t)b_{\mathrm{in}}(t_{1})} & =\bar{n}_{\mathrm{th}}\delta(t-t_{1}),\\
        \braket{b_{\mathrm{in}}(t)b_{\mathrm{in}}^{\dagger}(t_{1})} & =(1+\bar{n}_{\mathrm{th}})\delta(t-t_{1}),
    \end{aligned}
	\end{align}
	where $\bar{n}_\mathrm{th}=1/[\exp(\hbar\nu/k_BT_{\mathrm{vib}})]-1]$ is the average thermal photon numbers at the effective temperature $T_{\mathrm{vib}}$. For brevity, let us assume that the temperature
	satisfies $k_BT_{\mathrm{vib}}/\hbar\nu\ll 1$. According to the quantum regression theorem \cite{book_QuantumNoise,Book_carmichael_statistical_1999}, the non-zero correlation for the boson operator $b$ at $\tau>0$ is then obtained
	\begin{equation}
	    \braket{b(t)b^\dagger(t+\tau)}=e^{-(i\nu+\gamma)\tau}\label{eq:bb_Corr}.
	\end{equation}
	Introducing the momentum operator $p=i(b^\dagger-b)/\sqrt{2}$, one can get its correlation function
	\begin{equation}
	    \braket{p(t+\tau)p(t)}=\frac{1}{2}e^{-(i\nu+\gamma)\tau}\label{eq:pp_2t},
	\end{equation}
	and the variance $\braket{p^2}(t)=1/2$. Now, we can get the two-time correlation function for displacement operator $\mathcal{D}=\exp[\lambda(b-b^{\dagger})]=\exp(i\sqrt{2}\lambda p)$ as\begin{subequations}
	\begin{equation}
	    \begin{aligned}&\braket{\mathcal{D}(t+\tau)\mathcal{D}^{\dagger}(t)} \\=& \braket{e^{i\sqrt{2}\lambda p(t+\tau)}e^{-i\sqrt{2}\lambda p(t)}},\\
         =&\braket{e^{i\sqrt{2}\lambda[p(t+\tau)-p(t)]}}e^{\lambda^{2}[p(t+\tau),p(t)]},\\
         =&\braket{\sum_{k}\frac{(i\sqrt{2}\lambda)^{k}}{k!}[p(t+\tau)-p(t)]^{k}}e^{\lambda^{2}[p(t+\tau),p(t)]}.
        \end{aligned}\label{eq:DD_Corr1}
	\end{equation}
	Taking use of the Isserlis' theorem \cite{FloquetEngineering_Arxiv}, one can simplify Eq.~(\ref{eq:DD_Corr1}) into
	\begin{equation}
	    \begin{aligned}&\braket{\mathcal{D}(t+\tau)\mathcal{D}^{\dagger}(t)}\\= & e^{-\lambda^{2}\braket{[(p(t+\tau)-p(t)]^{2}}}e^{\lambda^{2}[p(t+\tau),p(t)]},\\
        = & e^{-\lambda^{2}[2\braket{p^{2}}-2\braket{p(t+\tau)p(t)}]},\\
        = & e^{-\lambda^{2}}e^{\lambda^{2}e^{-(i\nu+\gamma)(\tau)}}.
    \end{aligned}
	\end{equation}\end{subequations}

Similarly, we can also get the four-time correlation function for displacement operator at $t>t_1>t_2>t_3$,\begin{widetext}\begin{subequations}
        \begin{align}
            \braket{\mathcal{D}(t)\mathcal{D}^{\dagger}(t_{1})\mathcal{D}(t_{2})\mathcal{D}^{\dagger}(t_{3})}=&\braket{e^{i\sqrt{2}\lambda p(t)}e^{-i\sqrt{2}\lambda p(t_{1})}e^{i\sqrt{2}\lambda p(t_{2})}e^{-i\sqrt{2}\lambda p(t_{3})}},\nonumber\\
            =&e^{-\lambda^{2}\braket{[p(t)-p(t_{1})+p(t_{2})-p(t_{3})]^{2}}}e^{-\lambda^{2}[p(t)-p(t_{1}),p(t_{2})-p(t_{3})]}e^{\lambda^{2}[p(t),p(t_{1})]}e^{\lambda^{2}[p(t_{2}),p(t_{3})]}\nonumber,\\
            =&e^{-2\lambda^{2}[2\braket{p^{2}}-\braket{p(t)p(t_{1})}+\braket{p(t)p(t_{2})}-\braket{p(t)p(t_{3})}-\braket{p(t_{1})p(t_{2})}+\braket{p(t_{1})p(t_{3})}-\braket{p(t_{2})p(t_{3})}]}\label{eq:4T_DD}\\
            =&e^{-2\lambda^{2}}e^{2\lambda^{2}\braket{p(t)p(t_{1})}}e^{-2\lambda^{2}\braket{p(t)p(t_{2})}}e^{2\lambda^{2}\braket{p(t)p(t_{3})}}e^{2\lambda^{2}\braket{p(t_{1})p(t_{2})}}e^{-2\lambda^{2}\braket{p(t_{1})p(t_{3})}}e^{2\lambda^{2}\braket{p(t_{2})p(t_{3})}}.\nonumber
            \end{align}
            
Submitting Eq.~(\ref{eq:pp_2t}) into above equation, and taking the Taylor series expansion, we finally get the expression for the four-time correlation in the sum form
\begin{equation}
    \begin{aligned}
    &\braket{\mathcal{D}(t)\mathcal{D}^{\dagger}(t_{1})\mathcal{D}(t_{2})\mathcal{D}^{\dagger}(t_{3})}\\
    =&e^{-2\lambda^{2}}\sum_{\{k\}}\frac{(-1)^{k_{2}+k_{5}}\lambda^{2(k_{1}+k_{2}+k_{3}+k_{4}+k_{5}+k_{6})}}{\underset{j}{\prod}k_{j}!}e^{-(k_{1}+k_{2}+k_{3})(i\nu+\gamma)(t-t_{1})}e^{-(k_{2}+k_{3}+k_{4}+k_{5})(i\nu+\gamma)(t_{1}-t_{2})}e^{-(k_{3}+k_{5}+k_{6})(i\nu+\gamma)(t_{2}-t_{3})}
\end{aligned}
\end{equation}.
\end{subequations}
	
	\section{Dynamic of Electronic States \label{sec:Dynamic_DO}}
From the Langevin equation for the cavity field $a$ described by Eq.~(\ref{eq:Cavity Equation}), we can take the integration with respect of time to get the formal solution 
\begin{equation}
    a=\int_0^t dt_1\, e^{-(i\delta_c+\kappa)(t-t_1)}[-ig\sum_m\sigma_m(t_1)+\eta+(\sqrt{2\kappa_1}+\sqrt{2\kappa_2})a_\mathrm{in}(t_1)].\label{eq:Cavity_Int}
\end{equation}

By plugging Eq.~(\ref{eq:Cavity_Int}) into the formal solution of the dipole operator $\sigma_m(t)$ described by Eq.~(\ref{eq:Formula Solution}), and taking the averages, we find \begin{subequations}
\begin{align}
    \braket{\sigma_m}=&-ig\int_{0}^{t}dt_{1}e^{-(i\Delta+\Gamma_{\bot})(t-t_{1})}\int_{0}^{t_{1}}dt_{2}e^{-(i\Delta_{c}+\kappa)(t_{1}-t_{2})}\braket{\mathcal{D}_{m}(t)\mathcal{D}_{m}^{\dagger}(t_{1})[-ig\sum_{n\neq m}^{N}\sigma_{n}(t_{2})-ig\sigma_{m}(t_{2})+\eta]}.\label{eq:Formal1}
\end{align}
The item $\braket{\mathcal{D}_{m}(t)\mathcal{D}_{m}^{\dagger}(t_{1}\sigma_n(t)}$ reflects the induced interaction between $m$-th and $n$-th molecule via the quantized cavity field.   The item $\braket{\mathcal{D}_{m}(t)\mathcal{D}_{m}^{\dagger}(t_{1})\sigma_m(t_1)}$ relates to the modification of the frequency and decay rate.  When we considering the evolution for the Pauli operators, and submitting Eq.~(\ref{eq:Formula Solution}) into Eq.~(\ref{eq:Formal1}), the average value for Pauli operator is given by 
\begin{align}
    \braket{\sigma_m}=&ig^3\sum_{n\neq m}^N \int_{0}^{t}dt_{1}e^{-(i\Delta+\Gamma_{\bot})(t-t_{1})} \int_{0}^{t_{1}}dt_{2}e^{-(i\Delta_{c}+\kappa)(t_{1}-t_{2})} \int_{0}^{t_{2}}dt_3\,e^{-(i\Delta+\Gamma_{\bot})(t_{2}-t_{3})} \braket{\mathcal{D}_{m}(t)\mathcal{D}_{m}^{\dagger}(t_{1})\mathcal{D}_{n}(t_{2})\mathcal{D}_{n}^{\dagger}(t_{3})a(t_{3})}\nonumber\\
    &+ig^3\int_{0}^{t}dt_{1}e^{-(i\Delta+\Gamma_{\bot})(t-t_{1})} \int_{0}^{t_{1}}dt_{2}e^{-(i\Delta_{c}+\kappa)(t_{1}-t_{2})} \int_{0}^{t_{2}}dt_3\,e^{-(i\Delta+\Gamma_{\bot})(t_{2}-t_{3})}
    \braket{\mathcal{D}_{m}(t)\mathcal{D}_{m}^{\dagger}(t_{1})\mathcal{D}_{m}(t_{2})\mathcal{D}_{m}^{\dagger}(t_{3})a(t_{3})}\nonumber\\
    &-\eta g\int_{0}^{t}dt_{1}e^{-(i\Delta+\Gamma_{\bot})(t-t_{1})}\mathcal{D}_{m}(t)\mathcal{D}_{m}^{\dagger}(t_{1})\int_{0}^{t_{1}}dt_{2}e^{-(i\Delta_{c}+\kappa)(t_{1}-t_{2})}.\label{eq:Formal2}
\end{align}
\end{subequations} 

In the main text, we have roughly take the Markov approximation (under the  large vibrational relaxation condition $\gamma\gg\kappa,\gamma$) to separate the degrees of freedom for cavity field  and vibrational motion as
\begin{subequations}
\begin{equation}
\begin{aligned}
 \braket{\mathcal{D}_{m}(t)\mathcal{D}_{m}^{\dagger}(t_{1})a(t_1)}&\approx\braket{\mathcal{D}_{m}(t)\mathcal{D}_{m}^{\dagger}(t_{1})}\braket{a(t_1)},
\end{aligned}
\end{equation} and also the degrees of freedom for Pauli operator  and vibrational motion as
\begin{equation}
    \braket{\mathcal{D}_{m}(t)\mathcal{D}_{m}^{\dagger}(t_{1})\sigma_n(t_1)}\approx\braket{\mathcal{D}_{m}(t)\mathcal{D}_{m}^{\dagger}(t_{1})}\braket{\sigma_n(t_1)}.
\end{equation}
\end{subequations}
 This approach implies that we have roughly take the following approximation 
 \begin{equation}
     \braket{\mathcal{D}_m(t)\mathcal{D}_m^{\dagger}(t_{1})\mathcal{D}_m(t_{2})\mathcal{D}_m^{\dagger}(t_{3})}\approx \braket{\mathcal{D}_m(t)\mathcal{D}_m^{\dagger}(t_{1})}\braket{\mathcal{D}_m(t_{2})\mathcal{D}_m^{\dagger}(t_{3})}.
 \end{equation}
 In this case, all the displacement operators refer to the same vibrational mode. To test the feasibility of this approximation, we make a further study. According to the expression for the four-time correlation of the displacement operator in Eq.~(\ref{eq:4T_DD}), we can take the Laplace transformation for Eq.~(\ref{eq:Formal2}) to obtain
 \begin{equation}
     \braket{\bar{\sigma}_{m}}=ig^{3}\sum_{n\neq m}\frac{\mathcal{\mathcal{\bar{F}}}_{m}\mathcal{\mathcal{\bar{F}}}_{n}}{i\Delta_{c}+\kappa}\braket{\bar{a}}+ig^{3}\bar{\mathcal{F}}_{2,m}\braket{\bar{a}}-i\eta g\frac{\mathcal{\mathcal{\bar{F}}}_{m}}{s[i\Delta_{c}+\kappa]},
 \end{equation}
 where $\mathcal{\bar{F}}_n=\mathcal{\bar{F}}_n$ for identical molecules, and 
 \begin{equation}
     \begin{aligned}
     \bar{\mathcal{F}}_{2,m}=&\sum_{\{k\}}\frac{1}{\underset{j}{\prod}k_{j}!}e^{-2\lambda^{2}}(-1)^{k_{2}+k_{5}}\lambda^{2(k_{1}+k_{2}+k_{3}+k_{4}+k_{5}+k_{6})}\\
     &\times\frac{1}{[s+i\Delta+i(k_{1}+k_{2}+k_{3})\nu+(k_{1}+k_{2}+k_{3})\gamma+\Gamma_{\perp}][\cdots(1)\cdots][s+i\Delta+i(k_{1}+k_{2}+k_{3})\nu+(k_{1}+k_{2}+k_{3})\gamma+\Gamma_{\perp}]}
     \end{aligned}
 \end{equation}
with $[\cdots(1)\cdots]=[s+i\Delta_{c}+i(k_{2}+k_{3}+k_{4}+k_{5})\nu+(k_{2}+k_{3}+k_{4}+k_{5})\gamma+\kappa]$.

Considering the Laplace form of the cavity field described
	by Eq.~(\ref{eq:Cavity Laplace}), and using the final value theorem, we finally get
	\begin{equation}
	    \begin{aligned}
	    \braket{a}_{\mathrm{ss}}^\prime	&=\frac{\eta\mathcal{F}_{c}^{0}(-1+Ng^{2}\mathcal{F}_{c}^{0}\mathcal{F}_{m}^{0})}{-1+Ng^{4}\mathcal{F}_{c}^{0}[\mathcal{F}_{2,m}^{0}+(N-1)\mathcal{F}_{c}^{0}(\mathcal{F}_{m}^{0})^{2}]},\\
\braket{\sigma_{m}}_{\mathrm{ss}}^\prime	&=-i\eta g\frac{g^{2}\mathcal{F}_{c}^{0}[\mathcal{F}_{2,m}^{0}+(N-1)\mathcal{F}_{c}^{0}(\mathcal{F}_{m}^{0})^{2}]-\mathcal{F}_{c}^{0}\mathcal{F}_{m}^{0}}{-1+Ng^{4}\mathcal{F}_{c}^{0}[\mathcal{F}_{2,m}^{0}+(N-1)\mathcal{F}_{c}^{0}(\mathcal{F}_{m}^{0})^{2}]}.\label{eq:SS_2rd}
	    \end{aligned}
	\end{equation}
	
	\begin{figure}[H]
		\begin{centering}
			\includegraphics[scale=0.40]{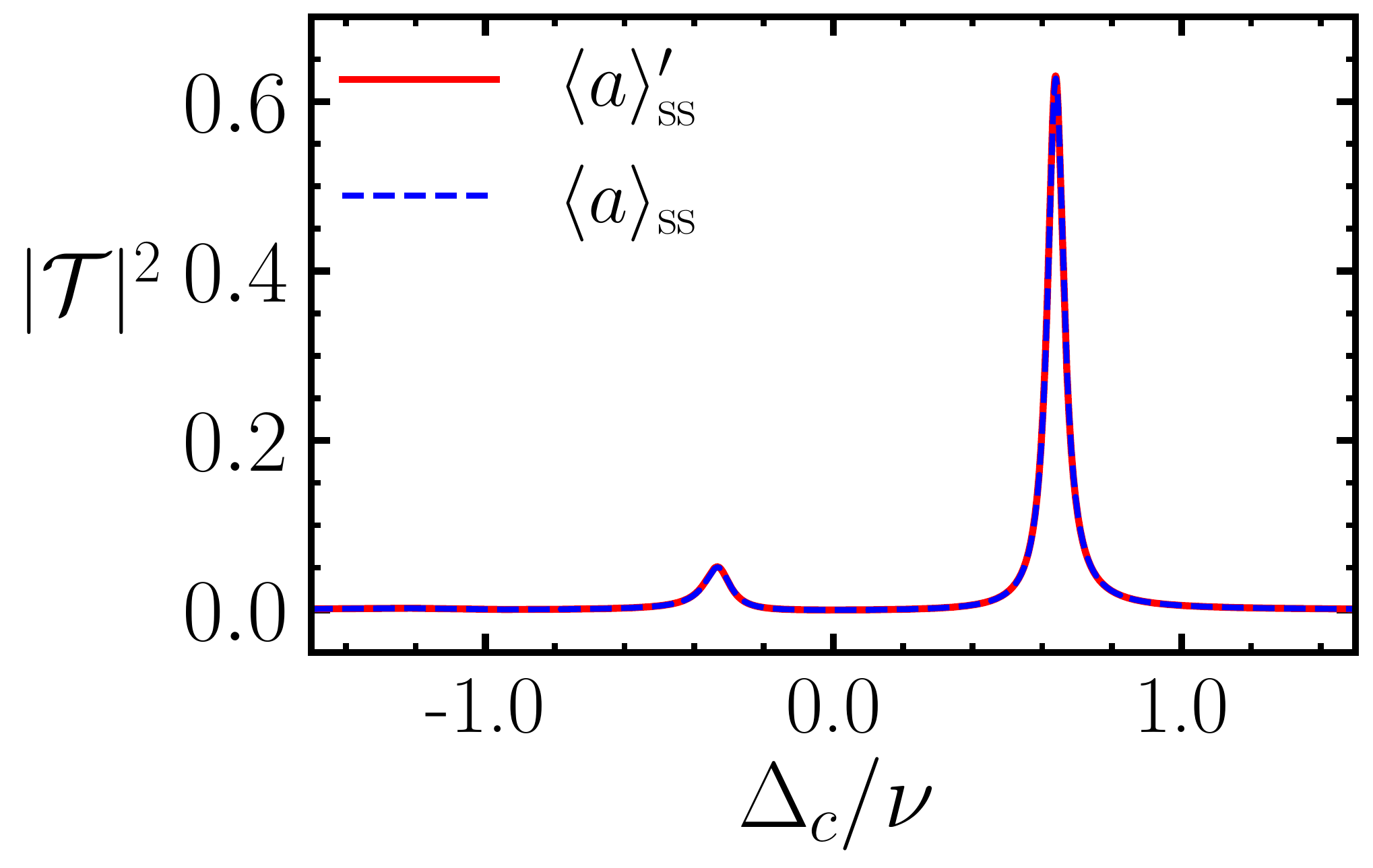}\caption{(Colored online) Cavity transmission at resonance
				 $\omega_{c}=\omega_{00}$  as a function
				of the detuning $\Delta_{c}/\nu$ with the number $N=200$  of organic
				molecules  at $\lambda=1$. The data is calculated  according to Eq.~(\ref{eq:SS_2rd}) (see red solid line) and Eq.~(\ref{eq:Cavity Steady}) (see blue dashed line), respectively, The other parameters are the same with
				Fig.~(\ref{fig:CavitySpec}). \label{fig:CavSpec2VS}}
			\par\end{centering}
	\end{figure}
	
To verify the validity of the Markov approximation to adiabatically eliminate the degree of freedom of vibrational motion,  we plot the  cavity transmission spectrum  at resonance $\omega_c=\omega_{00}$ with the number $N=200$  of organic molecules at $\lambda=1$. As illustrated by Fig.~\ref{fig:CavSpec2VS}, we can observe that same behaviors performed when we consider the four-time correlation (see red solid line) or just the two-time correlation (see blue dashed line).

	\section{Steady Population\label{sec: Appendix Steady State}}
	\subsection{Dynamics}
	From the Heisenberg-Langevin equation in Eq.~(\ref{eq:Langevin Equation_Form})
(or the master equation in Eq.~(\ref{eq:Master Equation}), we can derive
the dynamical equations for average number of photons in cavity $\braket{n_{c}}=\braket{a^{\dagger}a}$,
and the excited state population for $m-$th dye molecule $\braket{P_{m}}=\braket{\sigma_{m}^{\dagger}\sigma_{m}}$
	\begin{subequations}\begin{align}
			\frac{d}{dt}\braket{P_{m}}=&-\Gamma_{\Vert}\braket{P_{m}}-ig(\braket{a\sigma_{m}^{\dagger}}-\braket{a^{\dagger}\sigma_{m}}),\label{eq:Molecules_Pops} \\
			\frac{d}{dt}\braket{n_c}= & -2\kappa\braket{n_c}-ig\sum_{m}^{N}(\braket{a^{\dagger}\sigma_{m}}-\braket{a\sigma_{m}^{\dagger}})+\eta(\braket{a}+\braket{a^{\dagger}}),\label{eq:cavity-N}
			\end{align}\label{eq:Cav_Mol}\end{subequations}
		where $\Gamma_{\Vert}=2\Gamma$ is the longitudinal
		relaxation rate. By taking Laplace transformations for Eq.~(\ref{eq:Cav_Mol}),
		we will have the results\begin{subequations}\begin{align}
		s\braket{\bar{P}_m}= & -\Gamma_{\Vert}\braket{\bar{P}_m}-ig(\braket{\overline{a\sigma_{m}^{\dagger}}}-\braket{\overline{a^{\dagger}\sigma_{m}}}),\label{eq:Molecules_Pop_Laplace}\\
			s\braket{\bar{n}_c}= & -2\kappa\braket{\bar{n}_c}-ig\sum_{m}^{N}(\braket{\overline{a^{\dagger}\sigma_{m}}}-\braket{\overline{a\sigma_{m}^{\dagger}}})+\eta(\braket{\overline{a}}+\braket{\overline{a^{\dagger}}})\label{eq:Cavity_N_Laplace}.
			\end{align}\label{eq:Cavity_Molecules_Laplace}
		\end{subequations}
		
The presence of the operator $a^{\dagger}\sigma_{m}$ in Eq.~(\ref{eq:Cavity_Molecules_Laplace}),
relates to the nonlinear behavior of the HTC model. In the limit of
weak driving $\eta\ll\kappa$, the cavity field in steady situation
is approximately in vacuum. Under this situation, the noise correlation
between cavity field and the molecules as well as the noise correlation
between molecules will play important roles. Therefore, we have to
give the dynamics equations for the operator $a^{\dagger}\sigma_{m}$.
While considering the coupling between the electronic states and molecular
conformation, we start form the Langevin equation for the cavity filed
operator $a$ and the polarized operator $\tilde{\sigma}_{m}$ as
given in Eq.~(\ref{eq:Cavity-Pauli}),
to derive the dynamic equation for $a^{\dagger}\tilde{\sigma}_{m}$
\begin{subequations}
\begin{align}\frac{d}{dt}a^{\dagger}\tilde{\sigma}_{m}= & (\frac{d}{dt}a^{\dagger})\tilde{\sigma}_{m}+a^{\dagger}\frac{d}{dt}\tilde{\sigma}_{m}\label{eq:Pauli-Cavity Equation}\\
= & [i(\Delta_{c}-\Delta)-(\kappa+\Gamma_{\bot})]a^{\dagger}\tilde{\sigma}_{m}+ig\mathcal{D}_{m}^{\dagger}(\sigma_{m}^{\dagger}\sigma_{m}-a^{\dagger}a)+ig\mathcal{D}_{m}^{\dagger}\sum_{n\neq m}^{N}\sigma_{n}^{\dagger}\sigma_{m}+\sqrt{2\kappa}A_{in}^{\dagger}\tilde{\sigma}_{m}+\sqrt{2\Gamma_{\bot}}\mathcal{D}_{m}^{\dagger}a^{\dagger}\sigma_{m}^{\mathrm{in}}.\nonumber
\end{align}
Integrating with respect to the time $t$ on both sides of Eq.~(\ref{eq:Pauli-Cavity Equation}),
then the formal solution for the operator $a^{\dagger}\sigma_{m}$
will be obtained
\begin{equation}
a^{\dagger}\sigma_{m}=\int_{0}^{t}dt_{1}e^{[i(\Delta_{c}-\Delta)-(\kappa+\Gamma_{\bot})](t-t_{1})}\mathcal{D}_{m}(t)\mathcal{D}_{m}^{\dagger}(t_{1})\{ig[P_{m}(t_{1})-n_{c}(t_{1})]+\eta\sigma_{m}(t_{1})\}+\mathcal{X}_{m,\mathrm{ind}}^{t}+\mathcal{X}_{m,\mathrm{in}}^{t},\label{eq:Express-Cavity_Puali}
\end{equation}
with the dipole-dipole term
\begin{equation}
\mathcal{X}_{m,\mathrm{d}-\mathrm{d}}^{t}=ig\int_{0}^{t}dt_{1}e^{[i(\Delta_{c}-\Delta)-(\kappa+\Gamma_{\bot})](t-t_{1})}\mathcal{D}_{m}(t)\mathcal{D}_{m}^{\dagger}(t_{1})\sum_{n\neq m}^{N}\sigma_{n}^{\dagger}\sigma_{m}(t_{1}),\label{eq: Dipole-Dipole term}
\end{equation}
and the noise term 
\begin{equation}
\mathcal{X}_{\mathrm{in}}^{t}=\int_{0}^{t}dt_{1}e^{[i(\Delta_{c}-\Delta)-(\kappa+\Gamma_{\bot})](t-t_{1})}\mathcal{D}_{m}(t)\mathcal{D}_{m}^{\dagger}(t_{1})[\sqrt{2\kappa}a_{in}^{\dagger}\sigma_{m}(t_{1})+\sqrt{2\Gamma_{\bot}}\mathcal{D}_{m}^{\dagger}a^{\dagger}\sigma_{m}^{\mathrm{in}}(t_{1})].\label{eq: Noise term}
\end{equation}
\end{subequations}Since these individuals molecules are introduced
to couple a quantized cavity field, an efficient interaction between
molecules will be induced as described by Eq.~(\ref{eq: Dipole-Dipole term}).
To find the expression of the dipole-dipole term $\mathcal{X}_{m,\mathrm{d}-\mathrm{d}}^{t}$,
one need to consider the expression of $\sigma_{n}^{\dagger}\sigma_{m}$,
\begin{subequations} which can be derived through the following Langevin equation
\begin{equation}
\frac{d}{dt}\tilde{\sigma}_{n}^{\dagger}\tilde{\sigma}_{m}=-2\Gamma_{\bot}\tilde{\sigma}_{n}^{\dagger}\tilde{\sigma}_{m}+ig\mathcal{D}_{m}^{\dagger}\mathcal{D}_{n}(a^{\dagger}\sigma_{m}-a\sigma_{n}^{\dagger})+\sqrt{2\Gamma_{\bot}}\mathcal{D}_{m}^{\dagger}\mathcal{D}_{n}(\sigma_{n}^{\dagger}\sigma_{m}^{\mathrm{in}}+\sigma_{n}^{\mathrm{in}\dagger}\sigma_{m}).
\end{equation}
The evolution of $\sigma_{n}^{\dagger}\sigma_{m}(t)$ is then similarly
obtained as
\begin{equation}
\sigma_{n}^{\dagger}\sigma_{m}=ig\int_{0}^{t}dt_{1}e^{-2\Gamma_{\bot}(t-t_{1})}\mathcal{D}_{m}(t)\mathcal{D}_{m}^{\dagger}(t_{1})[a^{\dagger}\sigma_{m}(t_{1})-a\sigma_{n}^{\dagger}(t_{1})]\mathcal{D}_{n}(t_{1})\mathcal{D}_{n}^{\dagger}(t)+\mathcal{Y}_{\mathrm{in}}^{t},\label{eq:Evolution_DD}
\end{equation}
with the noise term
\begin{equation}
\mathcal{Y}_{\mathrm{in}}^{t}=\sqrt{2\Gamma_{\bot}}\int_{0}^{t}dt_{1}e^{-2\Gamma_{\bot}(t-t_{1})}\mathcal{D}_{m}(t)\mathcal{D}_{m}^{\dagger}(t_{1})[\sigma_{n}^{\dagger}\sigma_{m}^{\mathrm{in}}(t_{1})+\sigma_{n}^{\mathrm{in}\dagger}\sigma_{m}(t_{1})]\mathcal{D}_{n}(t_{1})\mathcal{D}_{n}^{\dagger}(t).
\end{equation}
\end{subequations}

From now on, the formal solution for the interactions between cavity and molecules characterized by Eq.~(\ref{eq:Express-Cavity_Puali}) as well as the interactions between molecules  depicted by Eq.~(\ref{eq:Evolution_DD}) have been obtained. To get the excited state population for total molecules, one can take the Markov approximation  to adiabatically eliminate the vibrational motion  without (or with) submitting Eq.~(\ref{eq:Evolution_DD}) into Eq.~(\ref{eq:Express-Cavity_Puali}), which relates to the perturbative treatment in the first (or second) order.

\subsection{Perturbative Treatment in the 1st Order}
In this subsection, let us take the perturbative treatment in the first order. Under the Markov approximation with the large vibrational
		relaxation, we can adiabatically eliminate the vibrational motion
		as introduced in Sec.~\ref{sec:Model} to get the dynamic equations
		for the item $\braket{a^{\dagger}\sigma_{m}}$ and $\braket{\sigma_{n}^{\dagger}\sigma_{m}}$
		in Laplace space\begin{subequations}
		\begin{align}
		    \braket{\overline{a^{\dagger}\sigma_{m}}}=&ig\bar{\mathcal{F}}_{m}^{\prime}(\braket{\bar{P}_m}-\braket{\bar{n}_c}+\sum_{n\neq m}^{N}\braket{\overline{\sigma_{n}^{\dagger}\sigma_{m}}})+\eta\bar{\mathcal{F}}_{m}^{\prime}\braket{\bar{\sigma}_{m}}\\		    			\braket{\overline{\sigma_{n}^{\dagger}\sigma_{m}}}=&ig\bar{\mathcal{F}}_{mn}(\braket{\overline{a^{\dagger}\sigma_{m}}}-\braket{\overline{a\sigma_{n}^{\dagger}}}),
		\end{align}
	\label{Pauli_Cavity&IntraMolec}\end{subequations}where $\bar{\mathcal{F}}_{m}^{\prime}$ is the
		Laplace form for $\mathcal{F}_{m}^{\prime}(t)=\exp[(i\Delta_{c}-i\Delta-\kappa-\Gamma_{\bot})t]\braket{\mathcal{D}_{m}(t)\mathcal{D}_{m}^{\dagger}(0)}$
		with the expression
		\begin{equation}
		\bar{\mathcal{F}}_{m}^{\prime}=\sum_{k}\frac{\lambda^{2k}}{k!}\frac{e^{-\lambda^{2}}}{s+i(\Delta-\Delta_{c}+k\nu)+(\kappa+\Gamma_{\bot}+k\gamma)},
		\end{equation}
		and $\bar{\mathcal{F}}_{mn}$ is the Laplace form for $\mathcal{F}_{mn}(t)=\exp(-2\Gamma_{\bot}t)\braket{\mathcal{D}_{m}(t)\mathcal{D}_{m}^{\dagger}(0)}\braket{\mathcal{D}_{n}(0)\mathcal{D}_{n}^{\dagger}(t)}$
		with the expression
		\begin{equation}
		\mathcal{\bar{F}}_{mn}=\sum_{k_{m},k_{n}}\frac{\lambda^{2(k_{m}+k_{n})}}{k_{m}!k_{n}!}\frac{e^{-2\lambda^{2}}}{s+i(k_{m}-k_{n})\nu+[2\Gamma_{\bot}+(k_{m}+k_{n})\gamma]}.
		\end{equation}
		
		For identical molecules, the Laplace forms for the population of the
		excited electric state $\braket{\overline{\sigma_{m}^{\dagger}\sigma_{m}}}$,
		the expect value for the interaction between cavity and molecule $\braket{\overline{a^{\dagger}\sigma_{m}}}$
		as well as the function $\bar{\mathcal{F}}_{m}^{\prime}$ are the
		same with that for any molecule ($m$), and the expect value for the
		intra-molecule interaction $\braket{\overline{\sigma_{n}^{\dagger}\sigma_{m}}}$,
		as well as the function $\mathcal{\bar{F}}_{mn}$ are the same for
		any molecule pair ($m$, $n$). As a result, we can simplify Eqs.~(\ref{eq:Cavity_Molecules_Laplace}),
		and (\ref{Pauli_Cavity&IntraMolec}), and transform them into matrix
		form 
		\begin{equation}
		\mathcal{M}_1\boldsymbol{V}_1+\eta\boldsymbol{V}_{1,\mathrm{in}}=0,
		\end{equation}
		with the drift matrix 
		\begin{equation}
		\mathcal{M}_1=\left(\begin{array}{ccccc}
		-2\kappa-s & iNg & -iNg & 0 & 0\\
		ig\bar{\mathcal{F}}_{m}^{\prime*} & -1 & 0 & -ig\bar{\mathcal{F}}_{m}^{\prime*} & -i(N-1)g\bar{\mathcal{F}}_{m}^{\prime*}\\
		-ig\bar{\mathcal{F}}_{m}^{\prime} & 0 & -1 & ig\bar{\mathcal{F}}_{m}^{\prime} & i(N-1)g\bar{\mathcal{F}}_{m}^{\prime}\\
		0 & -ig & ig & -\Gamma_{\Vert}-s & 0\\
		0 & -ig\bar{\mathcal{F}}_{mn} & ig\bar{\mathcal{F}}_{mn} & 0 & -1
		\end{array}\right),
		\end{equation} 
	the vector $\boldsymbol{V}_1=(\braket{\bar{n}_c},$
	$\braket{\overline{a\sigma_{m}^{\dagger}}},$ $\braket{\overline{a^{\dagger}\sigma_{m}}},$
	$\braket{\bar{P}_m},$ $\braket{\overline{\sigma_{m}^{\dagger}\sigma_{n}}})^{T}$
	and the input vector $\boldsymbol{V}_{1,\mathrm{in}}=(\braket{\overline{a}}+\braket{\overline{a^{\dagger}}},$
	$\bar{\mathcal{F}}_{m}^{\prime*}\braket{\bar{\sigma}_{m}^{\dagger}},$ $\bar{\mathcal{F}}_{m}^{\prime}\braket{\bar{\sigma}_{m}},$
	$0,$ $0)^{T}$. 

    We then obtain the solutions $\boldsymbol{V}_1=-\eta\mathcal{M}_1^{-1}\boldsymbol{V}_{1,\mathrm{in}}$
	by direct matrix inversion. According to the final value theorem, the excited state population
	for all molecules is finally given by
		\begin{equation}
	\mathcal{P}_{N}=\sum_{n=1}^{N}\braket{P_{m}}.
	\end{equation}

	If we set $N=1$, there will be only one molecule couples
	to the cavity and the inter-molecule interaction induced by cavity
	[depicted by Eq.~(\ref{eq: Dipole-Dipole term})] can be neglected.
	In such situation, the expect value for excited state population in
	steady state will be simplified as
	\begin{align}
	\mathcal{P}_{1} & =\frac{\eta g^{2}\mathfrak{R}\braket{a}_{\mathrm{ss}}\mathfrak{R}\mathcal{F}_{m}^{\prime0}-\eta g\kappa\mathfrak{I}\{\mathcal{F}_{m}^{\prime0}\braket{\sigma_{m}}_{\mathrm{ss}}\}}{\Gamma\kappa+g^{2}(\Gamma+\kappa)\mathfrak{R}\mathcal{F}_{m}^{\prime0}},\label{eq:PP1}
	\end{align}
	where $\mathcal{F}_m^{\prime0}=\underset{s\rightarrow0}{\lim}\bar{\mathcal{F}}_m^{\prime}$.
	
	Figure~\ref{fig:SSP} illustrates  the excited state population
	for all molecules  at resonance $\omega_c=\omega_{00}$ with  and  without  the  incoherent  coupling between  molecules. We compare the HTC model (red solid line) for varying number of molecule $N$, against the TC model (blue dashed line) in the limit $\alpha=0$ (a-d) and $\alpha=1$ (e-h). For finite $\lambda$,  the transfer rate of population from upper polaritonic states to lower polaritonic states and dark polaritonic states increased with the number of organic molecules increasing. Consequently the population  for upper polaritonic states will be reduced, and the population for dark polaritonic states is going to be increased. In addition, the transfer rate for the anti-Stokes process $\ket{\g}\ket{0}_\mathrm{vib}\rightarrow\ket{e}\ket{1}_\mathrm{vib}$ is enhanced with the increasing of the number $N$ of organic molecules. This reflects that the molecular conformation has been modified via the collective behavior.
	\begin{figure}[H]
		\begin{centering}
			\includegraphics[scale=0.3]{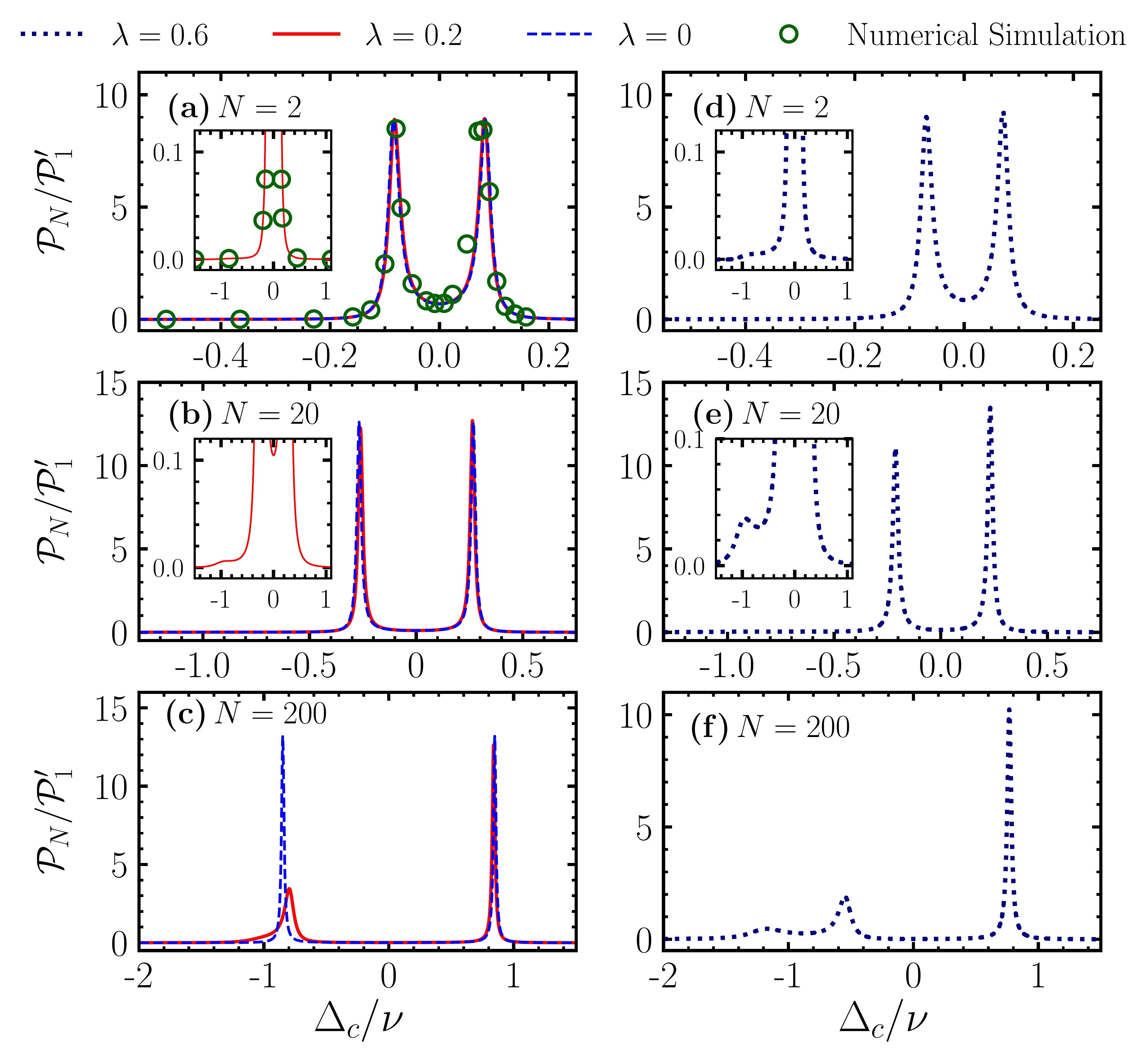}\caption{(Colored online) Steady population. The excited state population for
				all molecules at $\omega_{c}=\omega_{00}$ in steady state as a function
				of the detuning $\Delta_{c}/\nu$ with different number $N$ (increases as $2$, $20$ and $200$) of organic
				molecules  at $\lambda=0.6$ (red solid line)
				and $\lambda=0$ (blue dashed line) when (a-c) $\alpha=0$ and (d-f) $\alpha=1$. (a, d) The black dotted line shows the numerical
				simulation results for the case $\lambda=0.6$,  $N=2$ with the
				help of QuTiP. $P_{1}^{\prime}$ is the unit value given by Eq.~(\ref{eq:PP1})
				at $\Delta=0,$ $\lambda=0$. The other parameters are the same with
				Fig.~\ref{fig:CavitySpec}. \label{fig:SSP}}
			\par\end{centering}
	\end{figure}
 
\subsection{Perturbative Treatment in the 2nd Order}

To test the reasonableness of the approximation as previously mentioned, we will take perturbative treatment in higher order. Substituting Eq.~(\ref{eq:Evolution_DD}) into Eq.~(\ref{eq: Dipole-Dipole term}), and taking the averages, we find 
\begin{align}
\braket{\mathcal{X}_{m,\mathrm{d}-\mathrm{d}}^{t}}\approx & -g^{2}\sum_{n\neq m}^{N}\int_{0}^{t}dt_{1}\,e^{[i(\Delta_{c}-\Delta)-(\kappa+\Gamma_{\bot})](t-t_{1})}\int_{0}^{t_{1}}dt_{2}\,e^{-2\Gamma_{\bot}(t_{1}-t_{2})}\braket{\mathcal{D}_{m}(t)\mathcal{D}_{m}^{\dagger}(t_{2})}\braket{\mathcal{D}_{n}(t_{2})\mathcal{D}_{n}^{\dagger}(t_{1})}\nonumber \\
 & \times[\braket{a^{\dagger}\sigma_{m}}(t_{2})-\braket{a\sigma_{n}^{\dagger}}(t_{2})],\label{eq:DD Term Averange}
\end{align}
Here we have adiabatically eliminated the vibrational motion under
the large vibrational relaxation condition (i.e. $\gamma\gg\kappa$
and $\gamma\gg\Gamma)$. Noticing that the correlation function $\braket{\mathcal{D}_{m}(t)\mathcal{D}_{m}^{\dagger}(t_{2})}$
(and $\braket{\mathcal{D}_{m}(t)\mathcal{D}_{m}^{\dagger}(t_{1})}$)
depends on the time difference $t-t_{2}$ (and $t-t_{1}$) as described
in Eq.~(\ref{eq:DD CORR}), we can reformed Eq.~(\ref{eq:DD Term Averange})
as
\begin{align}
\braket{\mathcal{X}_{m,\mathrm{ind}}^{t}}= & -g^{2}e^{-\lambda^{2}}\sum_{n\neq m}^{N}\int_{0}^{t}dt_{1}\,e^{[i(\Delta_{c}-\Delta)-(\kappa+\Gamma_{\bot})](t-t_{1})}\int_{0}^{t_{1}}dt_{2}\,e^{-2\Gamma_{\bot}(t_{1}-t_{2})}e^{\lambda^{2}e^{-(\gamma+i\nu)(t-t_{2})}}\braket{\mathcal{D}_{n}(t_{2})\mathcal{D}_{n}^{\dagger}(t_{1})}\nonumber \\
 & \times[\braket{a^{\dagger}\sigma_{m}}(t_{2})-\braket{a\sigma_{n}^{\dagger}}(t_{2})].\nonumber \\
= & -g^{2}\sum_{n\neq m}^{N}\sum_{k}\int_{0}^{\infty}dt_{1}\int_{0}^{\infty}dt_{2}\,\mathcal{F}_{m}^{(k)}(t-t_{1})\mathcal{F}_{mn}^{(k)}(t_{1}-t_{2})[\braket{a^{\dagger}\sigma_{m}}(t_{2})-\braket{a\sigma_{n}^{\dagger}}(t_{2})],\label{eq:DD Convolution}
\end{align}
where $\tilde{\Delta}_k=\Delta_{c}-\Delta+k\nu$, $\tilde{\Gamma}_{1,k}=\kappa+\Gamma_{\bot}+k\gamma$, $\tilde{\Gamma}_{2,k}=2\Gamma_{\bot}+k\gamma$,  $\mathcal{F}_{m}^{(k)}(t)=\Theta(t)\lambda^{2k}\exp(-\lambda^{2})\exp[(i\tilde{\Delta}_k-\tilde{\Gamma}_{1,k})t]/{k!},$
and $\mathcal{F}_{mn}^{(k)}(t_{1}-t_{2})=\Theta(t_{1}-t_{2})\exp[-(ik\nu+\tilde{\Gamma}_{2,k})(t_{1}-t_{2})]\braket{\mathcal{D}_{n}(t_{2})\mathcal{D}_{n}^{\dagger}(t_{1})}.$
Since Eq.~(\ref{eq:DD Convolution}) represents a convolution, one can take use the Laplace transformation to rewritten Eq.~(\ref{eq:DD Term Averange}) into the Laplace space:
\begin{equation}
\braket{\bar{\mathcal{X}}_{m,\mathrm{ind}}^{t}}=-g^{2}\sum_{n\neq m}^{N}\mathcal{\bar{F}}_{mn}[\braket{\overline{a^{\dagger}\sigma_{m}}}-\braket{\overline{a\sigma_{n}^{\dagger}}}],
\end{equation}
where 
\begin{equation}
   \begin{aligned}\mathcal{\bar{F}}_{mn}= & \underset{k}{\sum}\mathcal{\bar{F}}_{m}^{(k)}\mathcal{\bar{F}}_{mn}^{(k)},\\
= & \sum_{k_{m},k_{n}}\frac{\lambda^{2(k_{m}+k_{n})}}{k_{m}!k_{n}!}\frac{e^{-2\lambda^{2}}}{[s+i(\Delta-\Delta_{c}+k_{m}\nu)+(\kappa+\Gamma_{\bot}+k_{m}\gamma)][s+i(k_{m}-k_{n})\nu+2\Gamma_{\bot}+(k_{m}+k_{n})\gamma]}.
\end{aligned}
\end{equation}
 Via substituting Eq.~(\ref{eq:Formula Solution}) into Eq.~(\ref{eq:Express-Cavity_Puali}), and adiabatically eliminating the degree of freedom of vibrational motion,  we similarly obtain the evolution of $\braket{a^{\dagger}\sigma_{m}}$ in Laplace space
\begin{equation}
    \braket{\overline{a^{\dagger}\sigma_{m}}}=ig[\bar{\mathcal{F}}_{m}^{\prime}(\braket{\overline{P}_{m}}-\braket{\bar{n}_{c}})-\eta\mathcal{\bar{F}}_{cm}^{\prime}\braket{\bar{a}}]-g^{2}\sum_{n\neq m}^{N}\mathcal{\bar{F}}_{mn}[\braket{\overline{a^{\dagger}\sigma_{m}}}-\braket{\overline{a\sigma_{n}^{\dagger}}}],\label{eq:Pauli-Cavity_Laplace}
\end{equation}
 where $\bar{\mathcal{F}}_{m}^{\prime}$ is the Laplace form for $\mathcal{F}_{m}^{\prime}(t)=\exp[(i\Delta_{c}-i\Delta-\kappa-\Gamma_{\bot})t]\braket{\mathcal{D}_{m}(t)\mathcal{D}_{m}^{\dagger}(0)}$
with the expression
\begin{equation}
\bar{\mathcal{F}}_{m}^{\prime}=\sum_{k}\frac{\lambda^{2k}}{k!}\frac{e^{-\lambda^{2}}}{s+i(\Delta-\Delta_{c}+k\nu)+(\kappa+\Gamma_{\bot}+k\gamma)},
\end{equation}
and 
\begin{equation}
    \mathcal{\bar{F}}_{cm}^{\prime}=\sum_{k}\frac{\lambda^{2k}}{k!}\frac{e^{-\lambda^{2}}}{[s+i(\Delta-\Delta_{c}+k\nu)+(\kappa+\Gamma_{\bot}+k\gamma)][s+i(\Delta+k\nu)+(\Gamma_{\bot}+k\gamma)]}.
\end{equation}

For identical molecules, the Laplace forms for the population of the
excited electric state $\braket{\overline{\sigma_{m}^{\dagger}\sigma_{m}}}$,
the expect value for the interaction between cavity and molecule $\braket{\overline{a^{\dagger}\sigma_{m}}}$
as well as the function $\bar{\mathcal{F}}_{m}^{\prime}$ are the
same with that for any molecule ($m$), and the expect value for the
intra-molecule interaction $\braket{\overline{\sigma_{n}^{\dagger}\sigma_{m}}}$,
as well as the function $\mathcal{\bar{F}}_{mn}$ are the same for
any molecule pair ($m$, $n$). As a result, we can simplify Eqs.~(\ref{eq:Cavity_Molecules_Laplace}), and (\ref{eq:Pauli-Cavity_Laplace})
into matrix form 
\begin{equation}
\mathcal{M}_2\boldsymbol{V}_2+\eta\boldsymbol{V}_{2,\mathrm{in}}=0,
\end{equation}
with the drift matrix 
\begin{equation}
\mathcal{M}_2=\left(\begin{array}{cccc}
-2\kappa-s & -iNg & iNg & 0\\
-ig\bar{\mathcal{F}}_{m}^{\prime} & -(N-1)g^{2}\mathcal{\bar{F}}_{mn}-1 & (N-1)g^{2}\mathcal{\bar{F}}_{mn} & ig\bar{\mathcal{F}}_{m}^{\prime}\\
ig\bar{\mathcal{F}}_{m}^{\prime*} & (N-1)g^{2}\mathcal{\bar{F}}_{mn}^{*} & -(N-1)g^{2}\bar{\mathcal{F}}_{mn}^{*}-1 & -ig\bar{\mathcal{F}}_{m}^{\prime*}\\
0 & ig & -ig & -\Gamma_{\Vert}-s
\end{array}\right),
\end{equation}
the vector $\boldsymbol{V}_2=(\braket{\bar{n}_{c}},$
$\braket{\overline{a^{\dagger}\sigma_{m}}},$ $\braket{\overline{a\sigma_{m}^{\dagger}}},$
$\braket{\bar{P}_{m}})^{T}$ and the input vector $\boldsymbol{V}_{2,\mathrm{in}}=(\braket{\bar{a}}+\braket{\overline{a^{\dagger}}}, -ig\bar{\mathcal{F}}_{cm}^{\prime}\braket{\bar{a}}, ig\bar{\mathcal{F}}_{m}^{\prime*}\braket{\overline{a^{\dagger}}}, 0)^{T}$. We then obtain the solutions $\boldsymbol{V}_2=-\eta\mathcal{M}_2^{-1}\boldsymbol{V}_{\mathrm{2,in}}$
by direct matrix inversion. Finally, the excited state population
for total molecules is given
	\begin{equation}
	\mathcal{P}^{\prime\prime}_{N}=\sum_{n=1}^{N}\braket{P_{m}}.
	\end{equation}

	If we set $N=1$, the expect value for excited state population in
	steady state will be simplified as
	\begin{align}
	\mathcal{P}^{\prime\prime}_{1} & =\eta g^2\frac{\mathfrak{R}\braket{a}_{\mathrm{ss}}\mathfrak{R}\mathcal{F}_{m}^{\prime0}+\kappa\mathfrak{R}\{\mathcal{F}_{cm}^{\prime0}\braket{a}_{\mathrm{ss}}\}}{\Gamma\kappa+g^{2}(\Gamma+\kappa)\mathfrak{R}\mathcal{F}_{m}^{\prime0}},\label{eq:PP1_Prime}
	\end{align}
	where $\mathcal{F}_m^{\prime0}=\underset{s\rightarrow0}{\lim}\bar{\mathcal{F}}_m^{\prime}$, and $\mathcal{F}_{cm}^{\prime0}=\underset{s\rightarrow0}{\lim}\bar{\mathcal{F}}_{cm}^{\prime}$.

	\begin{figure}[H]
		\begin{centering}
			\includegraphics[scale=0.40]{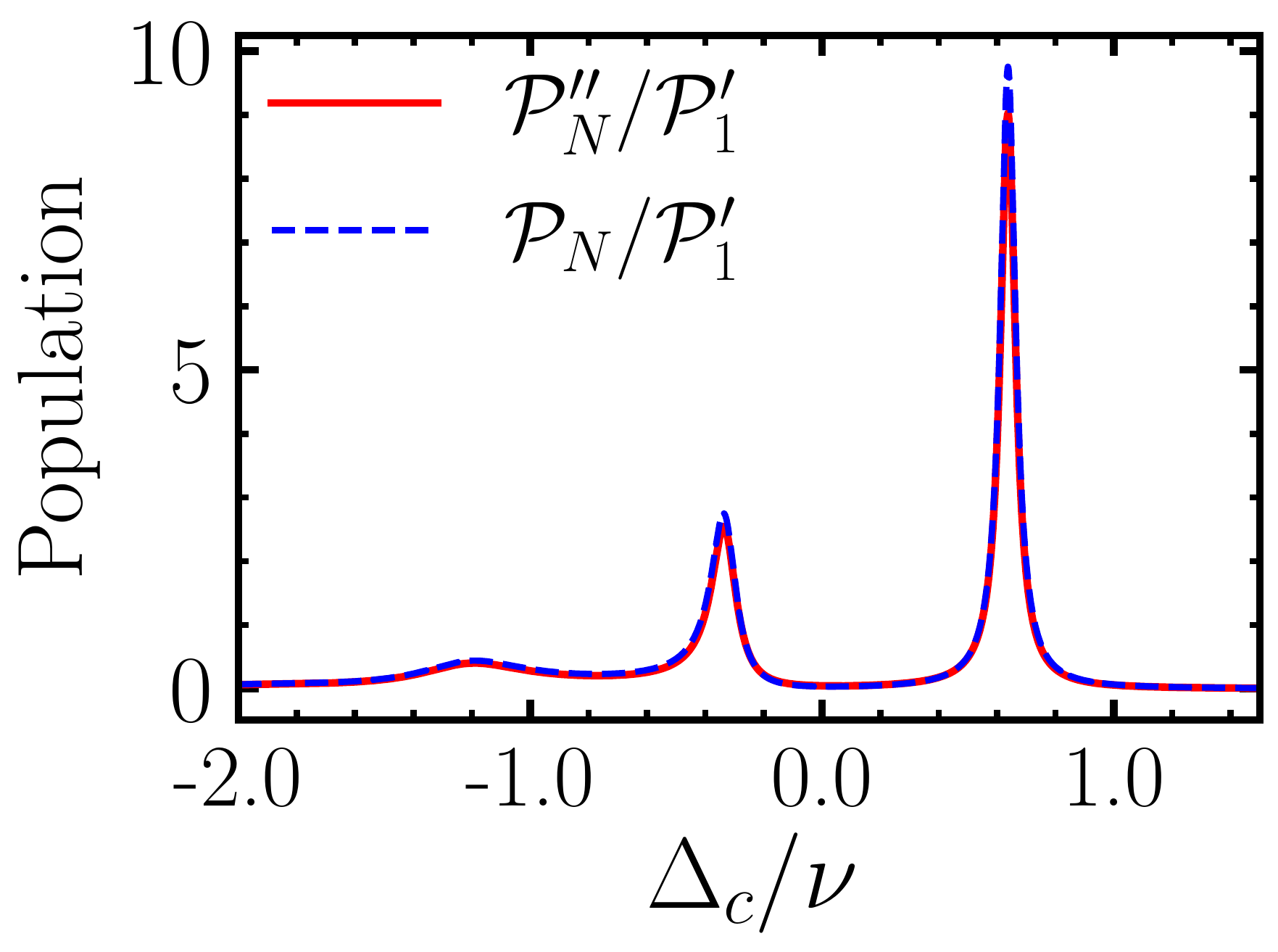}\caption{(Colored online) Steady population. The excited state population for
				total molecules at $\omega_{c}=\omega_{00}$ in steady state as a function
				of the detuning $\Delta_{c}/\nu$ with the  number $N=200$  of organic
				molecules  at $\lambda=1$. We compare the perturbative method in 1st order (blue dashed line) against the 2nd order (red solid line). $P_{1}^{\prime}$ is the unit value given by Eq.~(\ref{eq:PP1})
				at $\Delta=0,$ $\lambda=0$. The other parameters are the same with
				Fig.~\ref{fig:CavitySpec}. \label{fig:SSP_VS}}
			\par\end{centering}
	\end{figure}
	
	To test  the reasonableness of the Markov approximation to adiabatically eliminate the degree of freedom for the vibrational motion, we plot the excited steady population for total molecules various the detuning $\Delta_c/\nu$ at $\omega_{c}=\omega_{00}$ with the  number $N=200$  of organic molecules  at $\lambda=1$.   We can observe that similar behaviors performed when we take the perturbative method in second order (see red solid line) and first order (see blue dashed line).
	
	\end{widetext}
	
	\bibliographystyle{apsrev4-1}
	\bibliography{HTC}
	
\end{document}